\documentclass[a4paper,11pt]{article}
\usepackage{jinstpub}

\usepackage{
    graphicx,
    booktabs,
    textcomp,
    float,
    lineno,
    cleveref
}

\title{Convolutional L2LFlows: Generating Accurate Showers in Highly Granular Calorimeters Using Convolutional Normalizing Flows}

\author[a,b,1]{Thorsten Buss,\note{Corresponding author.}}
\author[b,c]{Frank Gaede,}
\author[a,c]{Gregor Kasieczka,}
\author[d,e]{Claudius Krause}
\author[f]{and David Shih}

\affiliation[a]{Institut f\"ur Experimentalphysik, Universit\"at Hamburg\\Luruper Chaussee~149, 22761 Hamburg, Germany}
\affiliation[b]{Deutsches Elektronen-Synchrotron DESY\\Notkestr. 85, 22607 Hamburg, Germany}
\affiliation[c]{Center for Data and Computing in Natural Sciences CDCS, Deutsches Elektronen-Synchrotron DESY\\Notkestr. 85, 22607 Hamburg, Germany}
\affiliation[d]{Institut f\"ur theoretische Physik, Universit\"at Heidelberg\\Philosophenweg 12, 69120 Heidelberg, Germany}
\affiliation[e]{Institut für Hochenergiephysik (HEPHY), \"Osterreichische Akademie der Wissenschaften (\"OAW)\\Dominikanerbastei 16, A-1010 Wien, \"Osterreich}
\affiliation[f]{NHETC, Department of Physics and Astronomy, Rutgers University\\Piscataway, NJ 08854, USA}

\emailAdd{Thorsten.Buss@uni-hamburg.de}


\arxivnumber{2405.20407}

\abstract{In the quest to build generative surrogate models as computationally efficient alternatives to rule-based simulations, the quality of the generated samples remains a crucial frontier. So far, normalizing flows have been among the models with the best fidelity. However, as the latent space in such models is required to have the same dimensionality as the data space, scaling up normalizing flows to high dimensional datasets is not straightforward. The prior L2LFlows approach successfully used a series of separate normalizing flows and sequence of conditioning steps to circumvent this problem. In this work, we extend L2LFlows to simulate showers with a 9-times larger profile in the lateral direction. To achieve this, we introduce convolutional layers and U-Net-type connections, move from masked autoregressive flows to coupling layers, and demonstrate the successful modelling of showers in the ILD Electromagnetic Calorimeter as well as Dataset~3 from the public CaloChallenge dataset.}

\begin{document}
\begin{flushright}
    HEPHY-ML-24-02
\end{flushright}
\maketitle
\flushbottom

\section{Introduction}

Motivated by the high computational cost of simulating particle interactions with complex detectors~\cite{HEPSoftwareFoundation:2017ggl,Boehnlein:2022}, a global effort is underway to develop reliable generative surrogate models based on machine learning.
To this end, network architectures based on generative adversarial networks (GANs)~\cite{Paganini:2017hrr,Paganini:2017dwg,deOliveira:2017rwa,Erdmann:2018kuh,Erdmann:2018jxd,Belayneh:2019vyx,Butter:2020qhk,ATLAS:2020,Carminati:2018khv,Musella:2018rdi,ATLAS:2022jhk,Ghosh:2020kkt,ATLAS:2021pzo,Hashemi:2023ruu,FaucciGiannelli:2023fow,Dogru:2024gpk}, 
variational autoencoders (VAEs)~\cite{Buhmann:2020pmy,Buhmann:2021lxj,Buhmann:2021caf,ATLAS:2022jhk,Cresswell:2022tof,Diefenbacher:2023prl,Hoque:2023zjt,Bieringer:2022cbs,Liu:2024kvv},
normalizing flows~\cite{Krause:2021ilc,Krause:2021wez,Schnake:2022,Krause:2022jna,Diefenbacher:2023vsw,Xu:2023xdc,Buckley:2023daw,Pang:2023wfx,Ernst:2023qvn,Schnake:2024mip,Du:2024gbp}, 
and various types of diffusion models~\cite{Mikuni:2022xry,Buhmann:2023bwk,Acosta:2023zik,Mikuni:2023tqg,Amram:2023onf,Buhmann:2023kdg,Buhmann:2023zgc,Jiang:2024ohg,Kobylianskii:2024ijw,Jiang:2024bwr,Favaro:2024rle}
have been explored (see~\cite{Hashemi:2023rgo} for a recent taxonomy).

Besides detector simulations, generative methods are used for various tasks in high-energy physics. These include end-to-end event generation~\cite{Hashemi:2019fkn,Otten:2019hhl,DiSipio:2019imz,Butter:2019cae,ArjonaMartinez:2019ahl,Alanazi:2020klf,Butter:2021csz,Butter:2023fov,Spinner:2024hjm,Vaselli:2024vrx}, phase space sampling~\cite{Bendavid:2017zhk,Klimek:2018mza,Chen:2020nfb,Gao:2020vdv,Bothmann:2020ywa,Gao:2020zvv,Danziger:2021eeg,Heimel:2022wyj,Janssen:2023ahv,Bothmann:2023siu,Heimel:2023ngj,Butter:2023ira}, parton showers~\cite{deOliveira:2017pjk,Andreassen:2018apy,Bothmann:2018trh,Dohi:2020eda,Kansal:2021cqp,Kach:2022qnf,Kach:2022uzq,Kansal:2022spb,Buhmann:2023pmh,Leigh:2023toe,Kach:2023rqw,Scham:2023usu,Scham:2023usu,Birk:2023efj,Li:2023xhj,Birk:2024knn,Mikuni:2024qsr}, hadronization~\cite{Ilten:2022jfm,Ghosh:2022zdz,Chan:2023ume,Bierlich:2023zzd,Chan:2023icm} and anomaly detection~\cite{Nachman:2020lpy,Andreassen:2020nkr,Stein:2020rou,Hallin:2021wme,Hallin:2022eoq,Raine:2022hht,Sengupta:2023xqy,Golling:2023yjq,Golling:2022nkl,Bickendorf:2023nej,Buhmann:2023acn,Sengupta:2023vtm,Krause:2023uww,Gandrakota:2024ruu}.

The fast simulation of particle showers in highly granular calorimeters, as under construction for the 
HL-LHC and envisaged for many future collider detectors, is a particularly worthwhile application of generative methods. Simulating such calorimeter showers is typically done with \textsc{Geant4}~\cite{geant4_2003,geant4_update,geant4_update_2016} a detailed simulation from first principles that is very demanding in terms of CPU resources due to the necessary creation and tracking of a huge number of secondary particles through the entire geometry for every shower. Highly granular calorimeters  sample particle showers across many layers, where every layer consists of an absorber material followed by a sensitive materials with readout cells of $O(1-10)~$mm sizes. The simulation of calorimeter showers need a substantial fraction of compute power at the LHC, even with less granular calorimeters that are currently in use, giving rise for large potential cost savings with generative methods.

While  normalizing flows achieve very high generative fidelity, they are so-far limited in their scalability to larger and more complex datasets.
The causal conditioning in \cite{Diefenbacher:2023vsw} was a first step in making flow-based architectures more scalable. 
At the core of the problem lies that normalizing flows learn a diffeomorphism between data space and latent space. 
While this --- in combination with tractable Jacobians --- allows the exact evaluation of the likelihood, it also means that the latent space and data space dimensionality are identical. 
As the number of trainable parameters in a fully connected architecture scales roughly with the product of the input and output dimensions of a layer, this would make data space volumes of e.g. 27k voxels prohibitively expensive.
The previous \textsc{L2LFlows} approach~\cite{Diefenbacher:2023vsw} circumvents this issue by splitting the data among one dimension and autoregressivly conditioning subsequent layers on the previous ones. This partially mitigates the problem, as at least in one dimension the scaling now becomes linear. 
Accordingly, \cite{Diefenbacher:2023vsw} demonstrated the learning of showers with 3k voxels.

However, the connections per calorimeter layer remained relatively naive.
In this work, we investigate three key innovations to improve the scalability and fidelity of normalizing flows:
i) convolutional layers assuming a approximate translational symmetry in lateral direction; ii) skip connections inspired by the U-net architecture~\cite{Ronneberger:2015}  to further improve the flow of information and network convergence; and iii) coupling layers instead of masked autoregressive flows for fast training and inference~\cite{Dinh:2014mzt,Dinh:2016pgf,Kingma:2018gyr,Durkan:2019nsq,Ernst:2023qvn} without the need of probability density distillation.

We demonstrate the feasibility of the proposed approach on the well established electromagnetic showers in the ILD first introduced in \cite{Buhmann:2020pmy} as well as on dataset~3 of the \textsc{CaloChallenge}~\cite{calochallenge}.

The rest of this paper is organized as follows: Section~\ref{sec:data} introduces the datasets used in this work. Section~\ref{sec:model} describes the model architecture. Section~\ref{sec:metrics} introduces the metrics used to evaluate the model. Section~\ref{sec:results} presents our results. Finally, section~\ref{sec:conclusion} concludes.

\section{Datasets}
\label{sec:data}
\begin{figure}
	\centering
	\includegraphics[height=0.2\textheight]{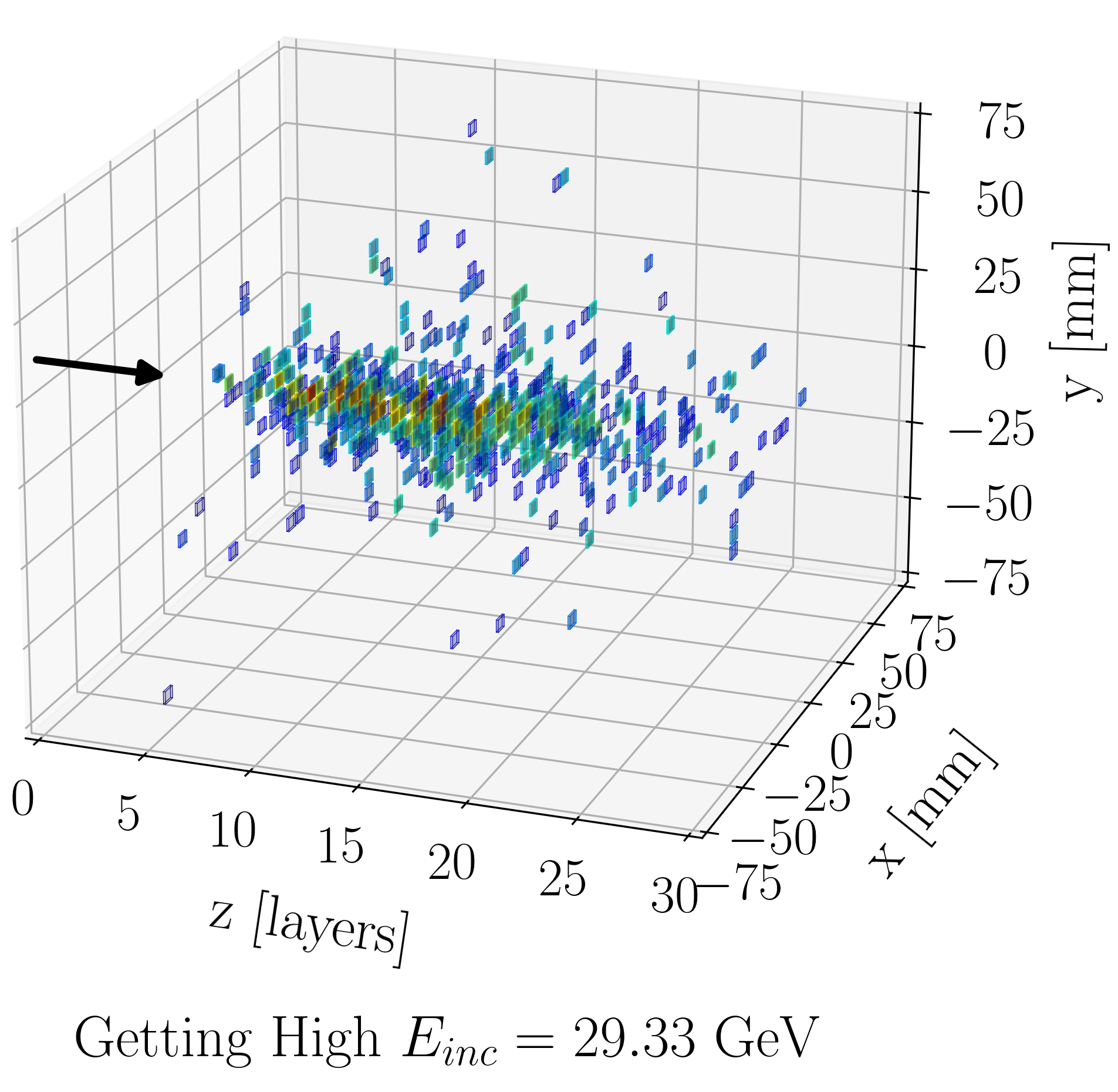}
    \hspace{0.04em}
	\includegraphics[height=0.2\textheight]{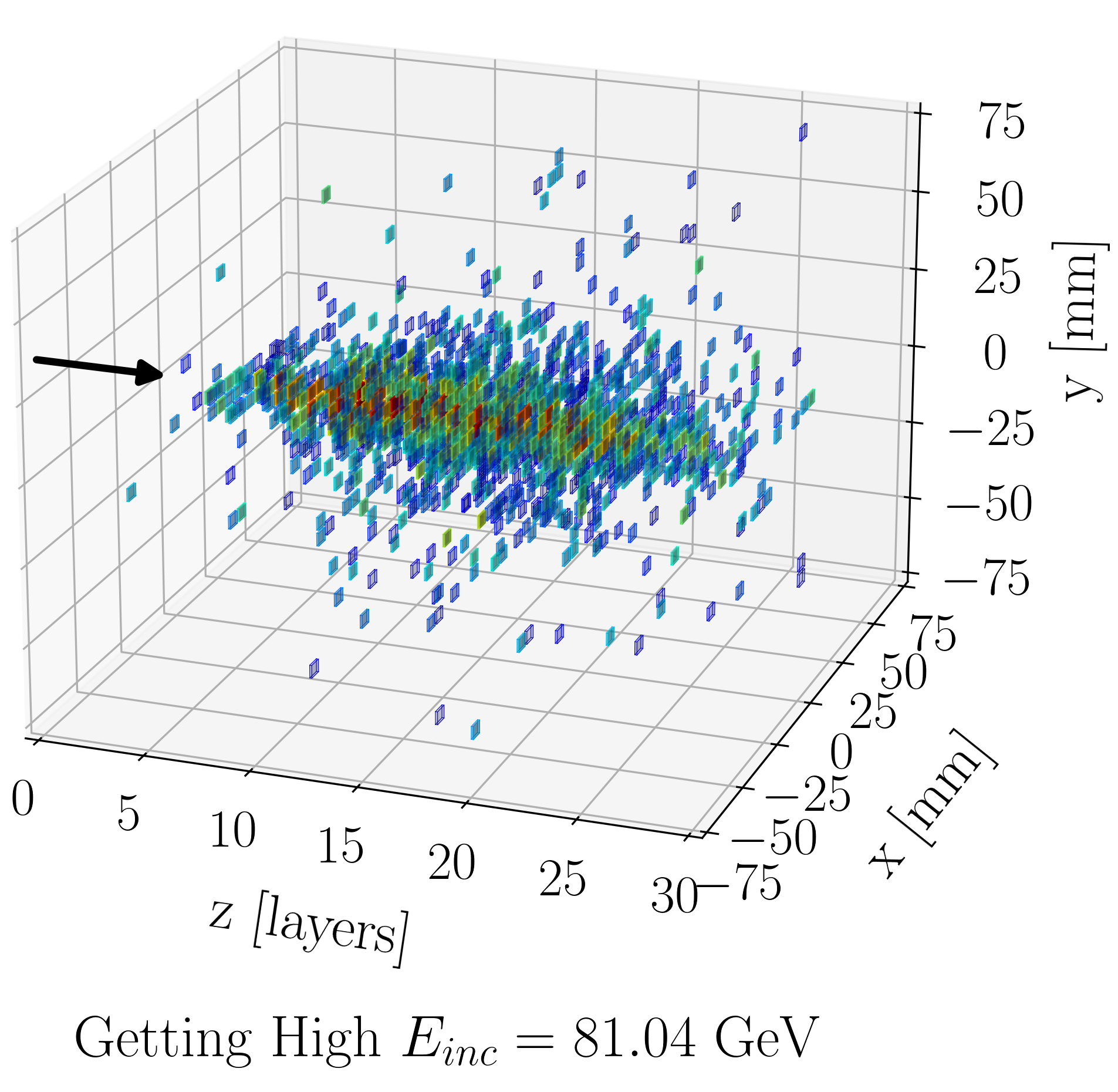}
    \hspace{0.04em}
	\includegraphics[height=0.2\textheight]{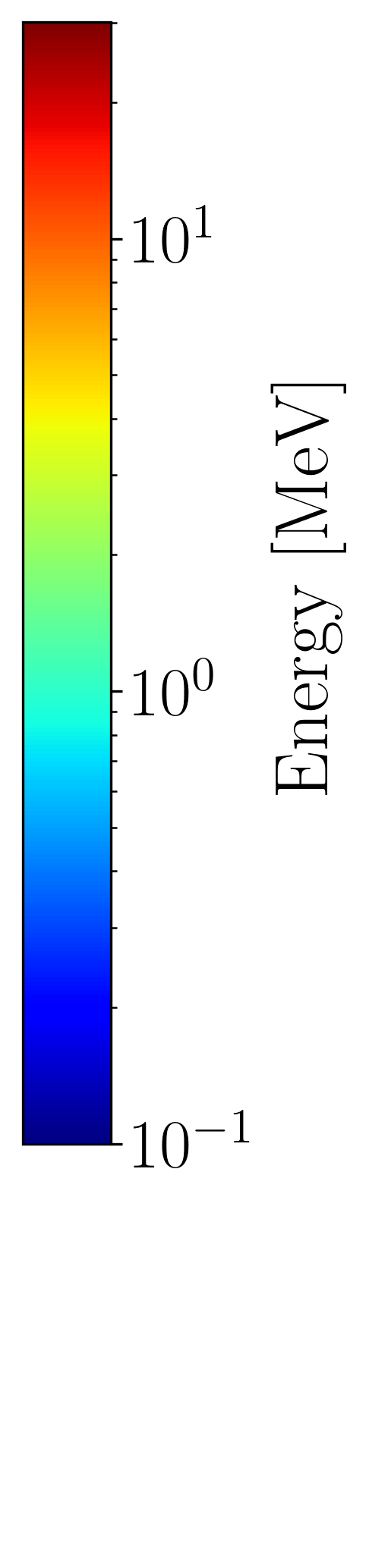}

	\includegraphics[height=0.2\textheight]{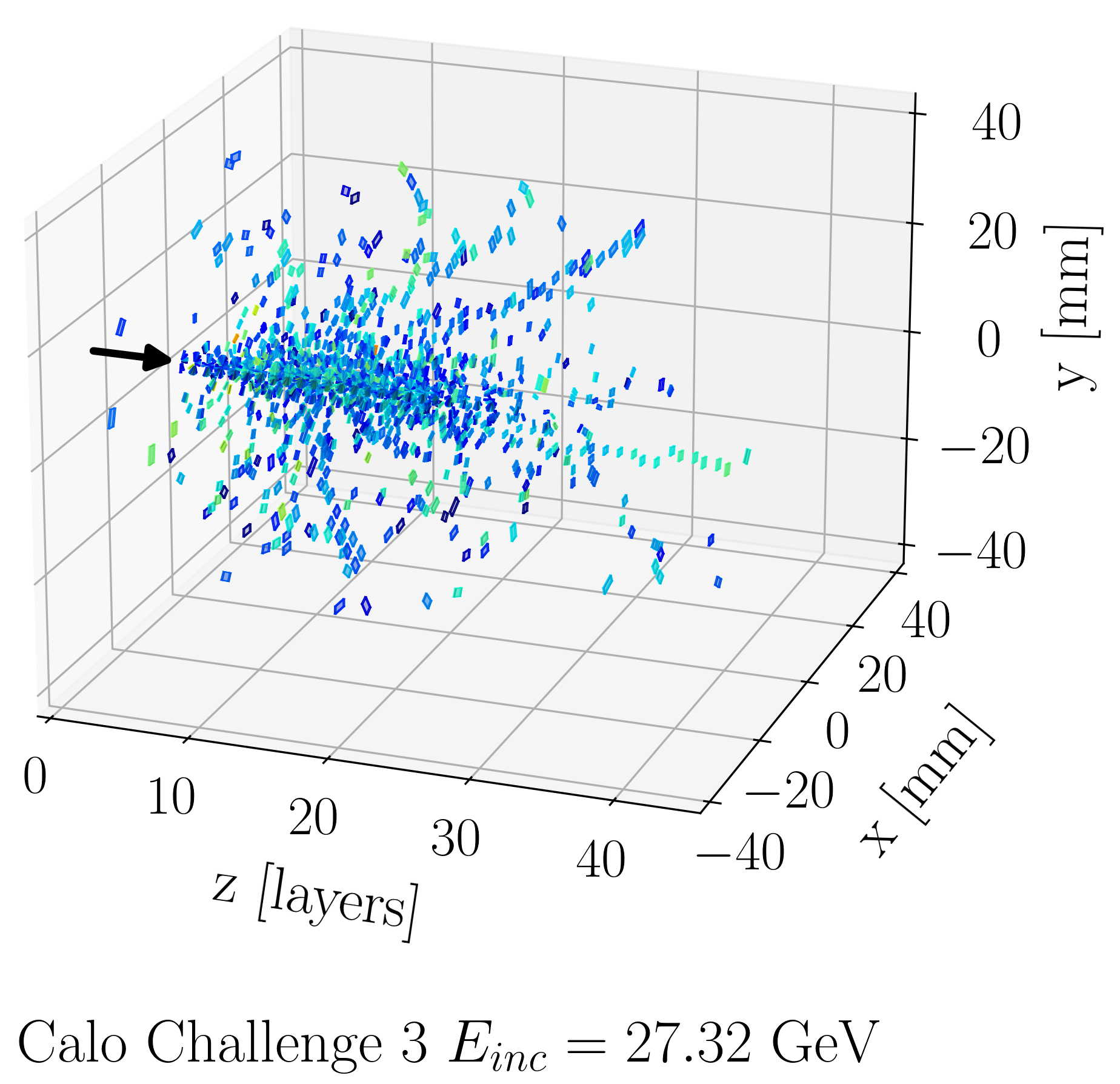}
	\includegraphics[height=0.2\textheight]{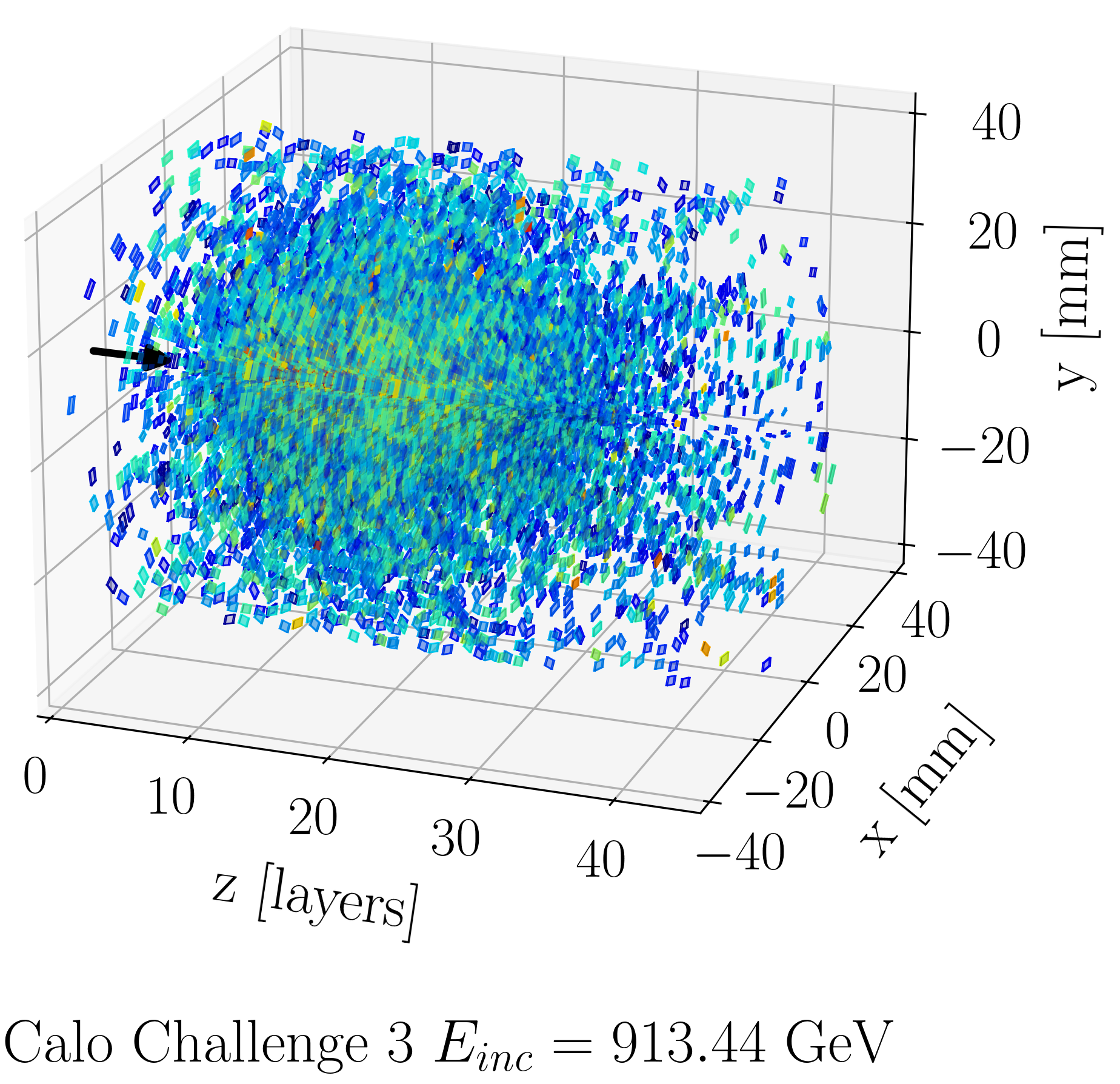}
	\includegraphics[height=0.2\textheight]{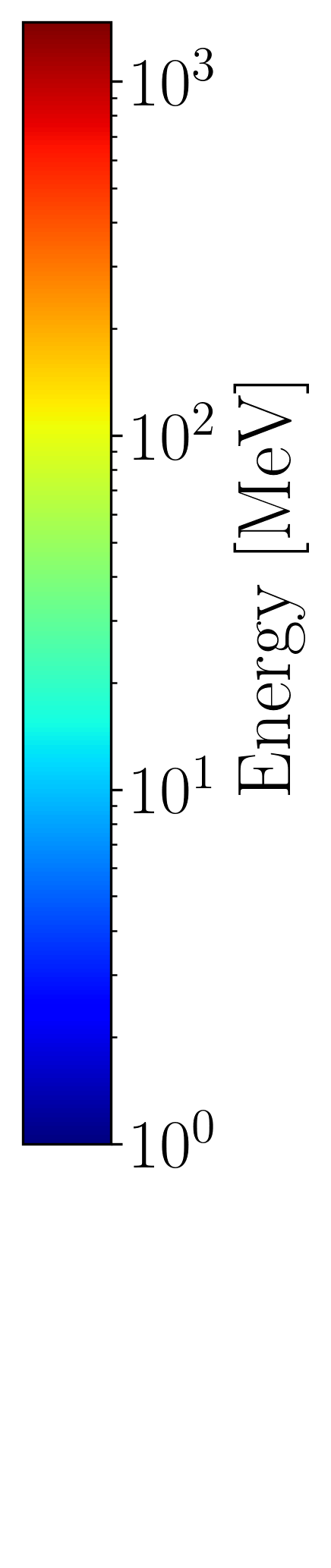}
	\caption{Visualization of single showers from the training data. The upper row shows examples from the 
    \textsc{GettingHigh} dataset, the lower row from \textsc{CaloChallenge} dataset~3. The left examples are for low incident particle energies, the right ones for high energies.
    }
    \label{fig:data}
\end{figure}

\subsection{GettingHigh}
For comparison with earlier work, we use the dataset generated for \cite{Buhmann:2020pmy}. It consists of photon showers in the electromagnetic calorimeter of the international large detector (ILD)~\cite{ILDConceptGroup:2020sfq}.

Two detectors are proposed for the international linear collider (ILC), one of which is the ILD. The ILD is tailored for Particle Flow~\cite{Thomson:2009rp}, an algorithm aiming to enhance detector resolution by reconstructing every individual particle. It combines precise tracking, good hermiticity, and highly granular calorimeters. The electromagnetic calorimeter (ECal) contains 30 silicon layers alternated with tungsten absorbers. It uses \(5\times5\;\text{mm}^2\) cells. The ILD detector's simulation, reconstruction, and analysis rely on the iLCSoft ecosystem~\cite{ilcsoft}, utilizing \textsc{Geant4}~\cite{geant4_2003,geant4_update,geant4_update_2016} and DD4hep~\cite{dd4hep} for realistic modeling.

The dataset consists of simulated photon showers with incident energies uniformly distributed from 10 to 100~GeV directed perpendicularly into the ECal. A staggered cell geometry results in small shifts between the layers, which generative models used to generate showers have to learn. A 30x30 cells bounding box around the incident point is chosen for the datasets, resulting in a data shape of 30x30x30. It has 950,000 samples, which we split into 765,000 training, 85,000 validation, and 100,000 test samples. Further details on the dataset are provided in~\cite{Buhmann:2020pmy}.

\subsection{Dataset~3 of the CaloChallenge}
To facilitate the development of deep generative models for calorimeter simulation, the Fast Calorimeter Simulation Challenge (\textsc{``CaloChallenge''})~\cite{calochallenge} was created. It consists of 4 calorimeter shower datasets of increasing dimensionality~\cite{CaloChallenge_ds1_v3,CaloChallenge_ds2,CaloChallenge_ds3}, all simulated with \textsc{Geant4}.

In this work, we work with the most complex of these datasets, dataset~3~\cite{CaloChallenge_ds3}. It consists of 100,000 electron showers, with log-uniform incident energies between 1~GeV and 1~TeV that were simulated with the Par04 example geometry~\cite{Par04}. This is an idealized calorimeter geometry made of 90 concentric cylinders of absorber (1.4\,mm of tungsten (W)) and active material (0.3\,mm of silicon (Si)). Its inner radius is 800\,mm and its depth is 153\,mm. All showers are perpendicular to the calorimeter surface, i.e. they are in the central $\eta=0$ slice. Particle entrance position and direction determine the origin and $z$-axis of a cylindrical coordinate system and showers are discretized into voxels, with boundaries parallel to these coordinate axes.  

For dataset~3 the voxels along $z$-axis corresponds to two physical layers (W-Si-W-Si) with a length of $\Delta z=3.4$\,mm ($=0.8 X_0$ of the absorber), yielding 45 read-out layers. The showers are segmented into radial ($r$) and angular ($\alpha$) bins. Dataset~3 has 18 radial segments and 50 azimuthal, for 900 voxels in each of the 45 layers (40500 total).

Each shower in the dataset consists of the total energy of the incident electron $E_{\rm inc}$ and the energy depositions recorded in each voxel $I_{ia}$, where $i$ is the layer index and $a$ is the voxel index within the layer. The minimum energy deposition in each voxel is 15.15~keV.

\section{Model}
\label{sec:model}
\begin{figure}
	\centering
	\includegraphics[width=0.7\textwidth]{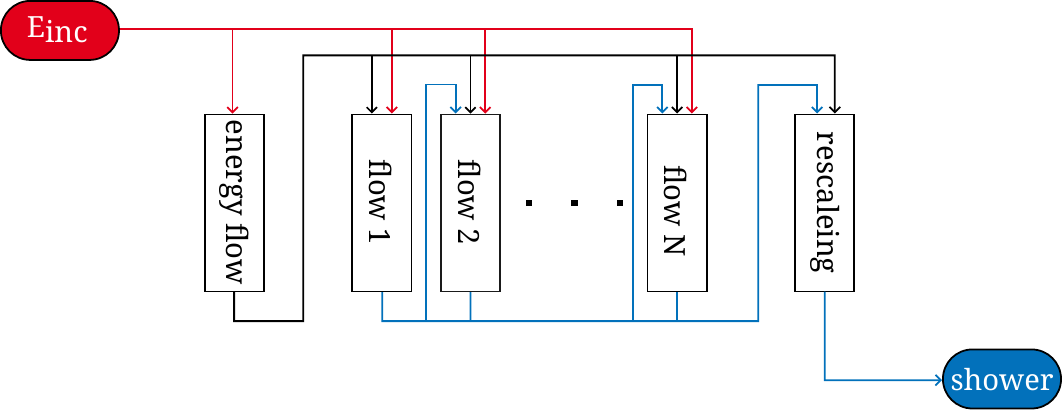}
	\caption{Diagram illustrating the overall architecture of \textsc{L2LFlows}. Arrows going into flows illustrate the conditional input of the flow. Arrows going out of flows illustrate what the flow has generated. The post processing is only applied to generated \textsc{GettingHigh} showers. }
    \label{fig:architecture}
\end{figure}

Our model is based on normalizing flows (NF)~\cite{Rezende:2015ocs,Kobyzev:2019ydm,Papamakarios:2019fms}. The idea of NFs is to learn a diffeomorphism \(f\) mapping a data distribution \(p(x)\) onto a simple prior distribution \(q(z)\) in latent space. The density in data space then is given by the change of variable formula,
\begin{equation}
	p(x)=q(f(x))\left|J_x\right|,
\end{equation}
where \(J_x\) denotes the Jacobian of \(f\) evaluated at position \(x\). This formula allows for likelihood-based training, where the loss function is the negative log-likelihood. To sample from an NF, one first draws samples from the prior distribution and then applies the inverse of \(f\) to these samples to map them back into data space. Networks used in NF setups have to be restricted in a way that they can only learn invertible mappings and that their Jacobians are tractable.

We aim to learn the distribution of calorimeter showers conditioned on the incident energy. As in previous approaches~\cite{Krause:2021ilc,Krause:2021wez,Diefenbacher:2023vsw,Buckley:2023daw}, we split this task into smaller pieces. An energy distribution flow learns the distribution of total deposited energies per calorimeter layer. For each calorimeter layer, a so-called causal flow learns the distribution of shower shapes in this particular layer. This results in \(N+1\) flows for a calorimeter with \(N\) layers.

To ensure consistency along the shower, later flows (in the direction of the shower evolution) are conditioned on the output of previous ones. The energy distribution flow is only conditioned on one global condition, the incident energy. The causal flows are conditioned on the incident energy, the layer energy for their layer, and the shower shape in up to five previous layers. Because the flows are conditioned on each other, during inference time, we have to draw samples from them sequentially. In this sense, it is an autoregressive model.

The original \textsc{L2LFlows} architecture~\cite{Diefenbacher:2023vsw} has two main drawbacks. First, it only achieves a relatively slow generation time. Second, it does not scale well to higher cardinalities. 
To overcome the first issue, we use coupling blocks~\cite{Dinh:2014mzt,Durkan:2019nsq} similar to ~\cite{Ernst:2023qvn} to speed up the generation time. Coupling-based flows can be evaluated in inverse direction almost as fast as in forward direction without requiring training a separate student model. To achieve better scalability, we used a convolution-based U-Net architecture~\cite{Ronneberger:2015} inside the coupling blocks. We will describe the architecture in more detail in the following sections.

\subsection{Energy Distribution Flow}
The energy distribution flow learns the distribution of layer energies, i.e., the sum of energy deposited per layer, conditioned on the incident energy. The density of this distribution can be expressed as
\begin{equation}
	p(E_1,\dots,E_N|E_{inc})
\end{equation}
where \(E_1\) to \(E_N\) denote the layer energies and \(E_{inc}\) denotes the incident energy. 
As this energy distribution flow is not the bottleneck of the generative task, we use the same 
preprocessing and architecture as in~\cite{Diefenbacher:2023vsw}.
It consists of 6 MADE blocks~\cite{Germain:2015yft,Papamakarios:2017tec,Kingma:2016wtg} with rational quadratic splines (RQS)~\cite{Gregory:1982,Durkan:2019nsq}. We apply permutation layers between these MADE blocks. A logit transformation is used to preprocess the layer energies, and a log transformation is used for the incident energy.

For \textsc{CaloChallenge} dataset~3, we observed outliers with too high energies being generated. To mitigate this issue, we rejected and redrew all samples with a ratio between deposited and incident energy above 2.6. This threshold was exceeded 22 times during the generation of 200,000 showers.

We use PyTorch~\cite{pytorch} and NFlows~\cite{nflows} to implement this model and train it using the ADAM optimizer~\cite{Kingma:2014}. 
By requiring the argument to the square root in the inversion of rational quadratic splines to be non-negative, we avoid previous numerical instabilities and can use single float instead of double float precision as was done in~\cite{Krause:2021ilc, Krause:2021wez,Krause:2022jna,Diefenbacher:2023vsw}.

\subsection{Causal Flows}
\begin{figure}
	\centering
	\includegraphics[height=4.5cm]{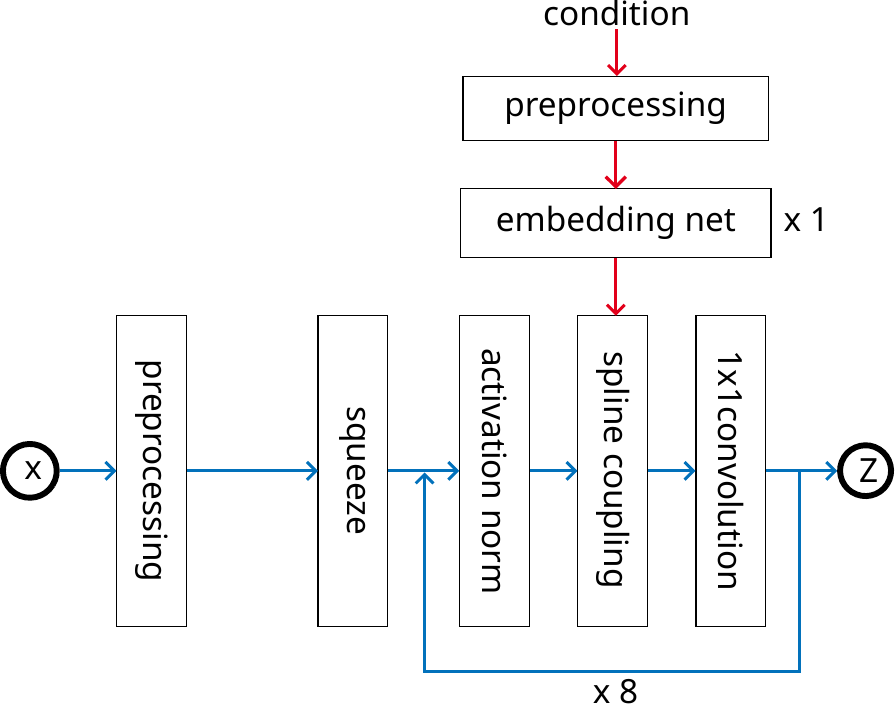}
	\hfil
	\includegraphics[height=4.5cm]{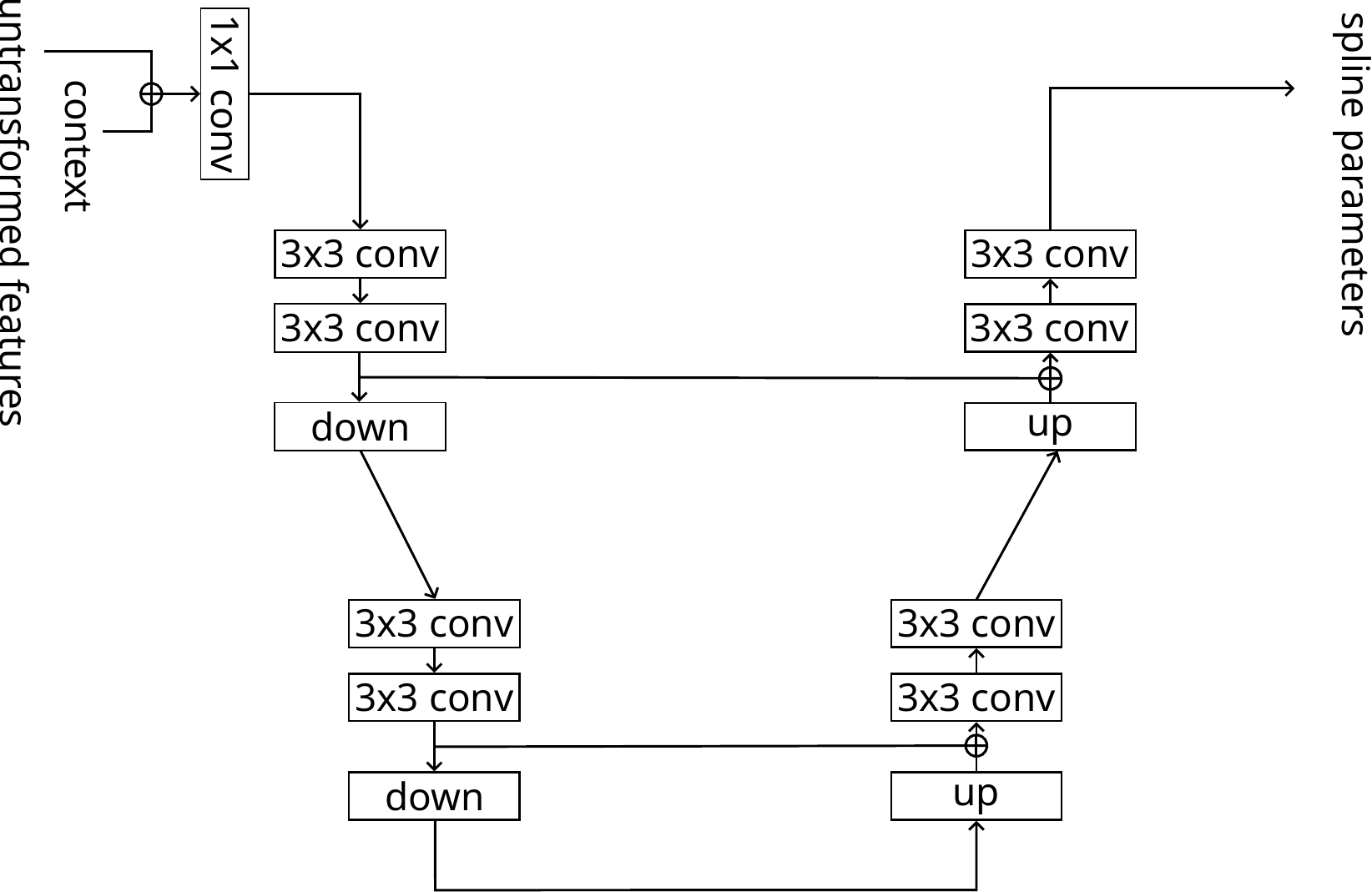}
	\caption{Diagrams illustrating the structure of a causal flow. Left: The overall structure of one single causal flow. Right: The U-Net used in the coupling blocks.}
    \label{fig:diagram}
\end{figure}
The task of the causal flows is to learn
\begin{equation}
	p(I_1,\dots,I_N|E_1,\dots,E_N,E_{inc})
\end{equation}
where \(I_i\in\mathbb{R}^{n\times n}\) is the shower shape in layer \(i\) given by the deposited energy in each calorimeter cell. As in the original \textsc{L2LFlows}, this task is split up into learning
\begin{equation}
	p(I_i|I_1,\dots,I_{i-1},E_1,\dots,E_N,E_{inc})
\end{equation}
for each layer \(i\) with an individual NF. By multiplying these conditional densities, one can gain the full joint distribution. As in the original \textsc{L2LFlows} \cite{Diefenbacher:2023vsw} approach, we assume approximate locality and only provide $I_j$ of up to five previous layers ($j \in \{i-5,i-4,i-3,i-2,i-1\})$, the energy deposited in layer \(i\), and the incident energy as condition to the flows. This helps the flow to focus on the most relevant information.

The architecture of the causal flows is shown in Figure~\ref{fig:diagram} on the left-hand side. Here, \(x\) is \(I_i\), the input to the shower, and \(z\) is the latent space vector.
Similar to CaloFlow~\cite{Krause:2021ilc}, we use log and logit transformations as preprocessing. Unlike earlier flow-based models for calorimeter simulations, we do not normalize the voxel energies to one but rather rescale with a constant value. This makes it easier for the flow to learn the voxel energy spectrum. Details on preprocessing can be found in~\ref{sec:preprocessing}.

In the following, we detail the additional improvements to the previous
\textsc{L2LFlows} architecture: 
i) replacing MADE block with coupling layers; 
ii) constructing convolutional sub-network based on the \textsc{U-Net} architecture.

As in~\cite{Ernst:2023qvn}, we use a coupling-based flow~\cite{Dinh:2014mzt,Durkan:2019nsq}. Similar to MADE blocks, spline coupling blocks use rational quadratic splines. Unlike MADE blocks, coupling blocks are not fully autoregressive. Instead, they split the input dimensions into two halves. The first half is fed into a sub-network that determines the spline parameters for the transformation of the second half. The first half is unchanged, allowing an inversion with a single network path, mitigating the need for a teacher-student approach~\cite{Krause:2021wez,Buckley:2023daw}. As for MADE blocks, the Jacobian is a triangular matrix allowing for a fast evaluation of the determinant.

In convolutional coupling flows such as in \cite{Dinh:2014mzt,Dinh:2016pgf,Kingma:2018gyr}, the information is usually split along the channel dimension to preserve spatial structures. 
To go from calorimeter data with a single channel to several channels, we apply a so-called squeezing operation~\cite{Dinh:2016pgf}. The idea is to stack pixels lying in small tiles into channels, trading in spatial dimensions for channel dimensions. To be able to concatenate input and condition, we need to bring the condition into the same shape. For this task, a small convolutional embedding network is used.

In flows like \textsc{RealNVP}~\cite{Dinh:2016pgf} or \textsc{GLOW}~\cite{Kingma:2018gyr}, a multi-scale architecture was used to learn local features first and then regional and global features of the data. This is achieved by iteratively applying further squeezing operations. We instead use only a single squeezing operation, leaving the task of learning features on different scales to U-Nets~\cite{Ronneberger:2015}. The U-Nets are deployed within the coupling blocks to calculate the spline parameters. We found that this makes the architecture more flexible and leads to better results.\footnote{U-Nets as sub-networks for a \textsc{GLOW}-like flow were also explored in~\cite{Park:2021}.}

Figure~\ref{fig:diagram} on the right-hand side shows an illustration of the U-Net used. The \(3\times{}3\) convolutions use same padding to preserve the data size. The down-sampling operations are convolutional layers with a stride equal to the kernel size, while the up-sampling operations are inverse convolutional layers. For the \textsc{CaloChallenge} dataset, we use cyclic padding in the angular dimension to respect the cyclic structure of the data.

The last two parts of the architecture are activation norms and 1x1 convolutions, both of which were introduced in \textsc{GLOW}. Since batch and layer normalization are not directly possible in flows, activation norms provide a viable alternative. The idea is to initialize the normalization on the first batch and then update it as model parameters. 1x1 convolutions replace the permutations often used in flows to ensure the information is split up differently in each coupling block. They are initialized as orthogonal transformations. A LU decomposition ensures fast invertibility; even so, the network is not restrained to orthogonal transformations~\cite{Kingma:2018gyr}.

\subsection{Training}
As the log-likelihood is additive under joining distributions, training each flow individually is equivalent to training all flows jointly. This makes training easier and allows for straightforward parallelization on several compute nodes.

Calorimeter data are sparse, meaning that most input values are zero. This poses a problem for training flows. Since the data points live in a submanifold of the data space, the density and therefore the loss are not bounded. This problem is usually solved by adding noise. As in the original CaloFlow, we add uniform noise between zero and \(1\)~keV to all voxels. Only adding uniform noise would lead to discontinuities in the density at zero and one keV. To avoid that, we fill zero voxels with log Gaussian distributed values before adding the uniform noise.

We trained one instance of the networks for 200 epochs on \textsc{GettingHigh} and one instance for 800 epochs on \textsc{CaloChallenge} dataset~3. The different amount of epochs is necessary because of the different training statistics. We applied a one-cycle learning rate scheduler~\cite{Smith:2018}, decoupled L2 regularization~\cite{Loshchilov:2017bsp}, and L2 gradient clipping. ADAM~\cite{Kingma:2014} was used as optimizer. As data augmentation, we rotated the \textsc{CaloChallenge} dataset~3 showers by random numbers of angular bins in the \(x-y\) plane. This is possible because the incident particles enter the calorimeter perpendicular to its surface and no magnetic field brakes the symmetry. Further details on architecture and training are provided in \ref{sec:architecture}.

\subsection{Postprocessing}
It is not guaranteed that the causal flows produces showers that match the layer energies generated by the energy distribution flow exactly. For that reason,\cite{Krause:2021ilc,Diefenbacher:2023vsw} rescaled showers per-layer to match the target. However, like the authors of~\cite{Buckley:2023daw}, we observe that rescaling generated \textsc{CaloChallenge} showers results in an excess of low energy hits due to zero voxels getting shifted above the threshold. 
Therefore, instead of applying postprocessing for \textsc{CaloChallenge}, the showers generated by the causal flows are used directly. For \textsc{GettingHigh} we use the same energy rescaling as in \cite{Diefenbacher:2023vsw}. It involves a way of identifying and masking empty voxels. See Appendix B.3 in \cite{Diefenbacher:2023vsw} for a detailed description.

We apply a further postprocessing step to the \textsc{GettingHigh} showers. This step maps the distribution of generated voxel energies onto the real voxel energy spectrum. To do so, it applies a one-dimensional function element-wise to the voxel energies. This function can be seen as a one-dimensional normalizing flow with the generated voxel energy spectrum as data distribution and the real voxel energy spectrum as latent distribution. Since we are in the one-dimensional case, this function can be evaluated numerically, and it is not necessary to learn it with a network. To speed up evaluations, we fitted a spline to the function. Further details can be found in~\ref{sec:postprocessing}.

\section{Metrics}
\label{sec:metrics}

\subsection{Observables}
\label{sec:observables_m}

Generative models aim to follow training data distributions as closely as possible. 
For applications in physics, it is especially interesting how accurately the distributions of certain high-level physics observables are reproduced. We give a brief description of all physics observables considered in the following.
In addition to showing histograms, we also report the Wasserstein distance~\cite{Kansal:2022spb} between the ground truth and generated samples.

\begin{table}[htbp]
    \centering
    \begin{tabular}{p{0.3\linewidth} p{0.6\linewidth}}
        \multicolumn{2}{c}{\bf Hit-level Observables} \\
        Voxel energy spectrum & hit energies regardless of their position \\
        Longitudinal profile & energy-weighted layer indexes \\
        Radial profile & energy-weighted distances from the incident point in the \(x\)-\(y\)-plane \\
        X profile & energy-weighted \(x\) positions \\ \\ 

        \multicolumn{2}{c}{\bf Shower-level Observables} \\
        Energy ratio & total measured energy summed over all voxels divided by the incident energy \\
        Occupancy & fraction of active voxels in a shower (Active voxels are all voxels with energies above the data-set-specific threshold.) \\
        Center of gravity & energy-weighted mean of \(x\), \(y\), or longitudinal positions evaluated on each shower individually \\
        Shower width & energy-weighted standard deviation of \(x\), \(y\), or longitudinal positions evaluated on each shower individually \\
    \end{tabular}
    \label{tab:observables}
\end{table}

\subsection{Classifier}
\label{sec:classifier_m}
Histograms and one-dimensional Wasserstein distances are useful to test how well one-dimensional distributions match, but they do not account for correlations.  Instead, we use classifier tests to compare generative models based on higher-dimensional distributions~\cite{Lopezpaz:2018,Krause:2021ilc,Krause:2021wez,Buckley:2023daw,Diefenbacher:2023vsw}.

We train a high-level classifier on nine observables to distinguish ML-generated and original simulation data. These are the eight shower-level observables described in the previous section and the incident energy. The incident energy allows the classifier to detect mismodelings of correlations between the incident energy, as well as the shower observables. Additionally, a classifier is trained on the hit-level information, using a convolutional architecture. Further details on the classifiers can be found in \ref{sec:classifier_a}.

As the authors of~\cite{Krause:2021ilc,Krause:2021wez,Buckley:2023daw}, we use the area under the receiver operating characteristic curve (AUC) and the Jensen–Shannon divergence (JSD) as performance scores.

\section{Results}
\label{sec:results}
\begin{figure}
    \centering
    \includegraphics[width=0.9\linewidth]{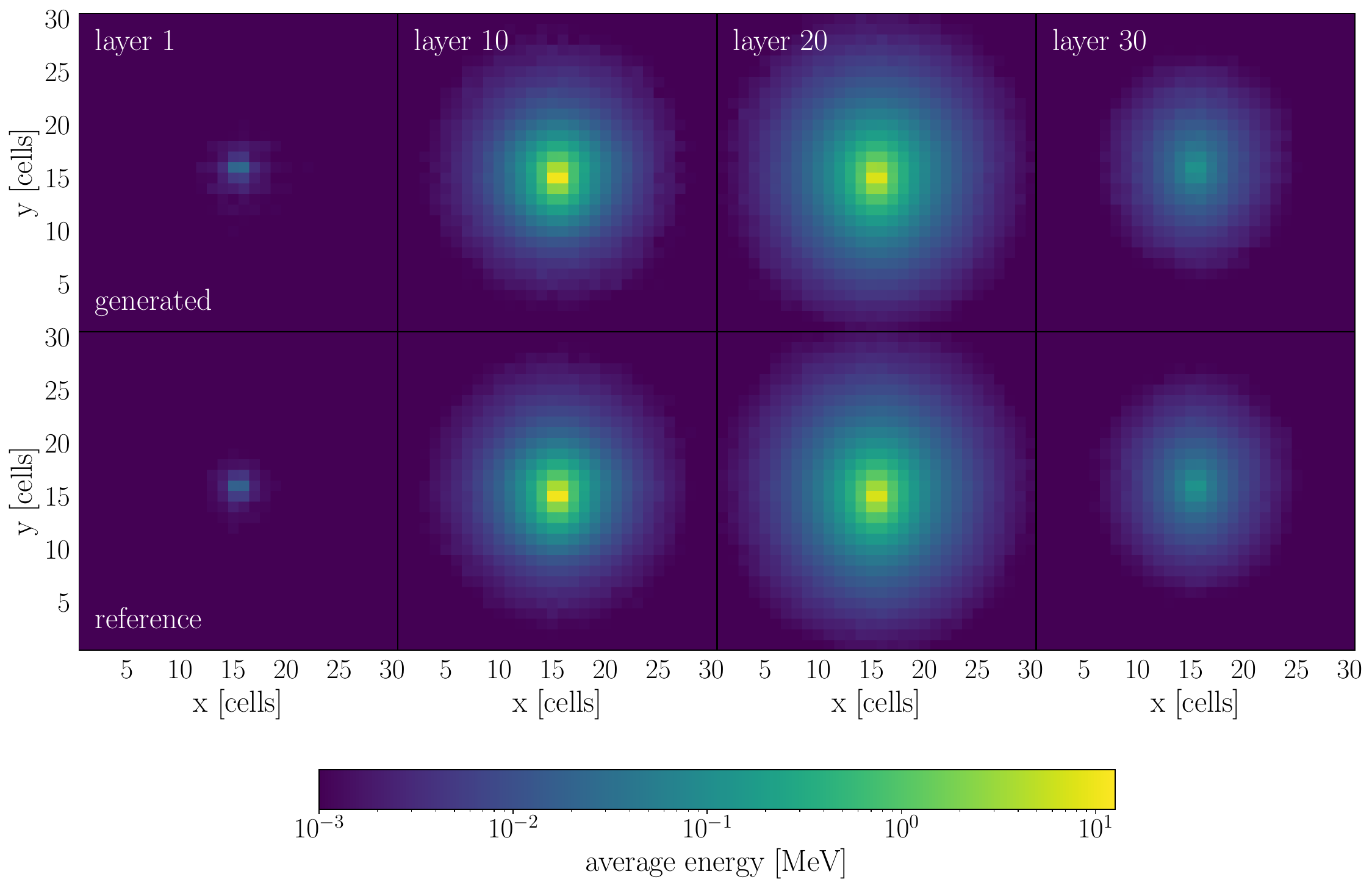}
    \caption{Average deposited energy for all cells in layer 1, 10, 20 and 30 overall flow generated (upper row) and \textsc{Geant4} simulated (lower row) sample. Shown are the results for \textsc{GettingHigh}.}
    \label{fig:overlay_gh}
    \vspace{2em}
    \includegraphics[width=0.9\linewidth]{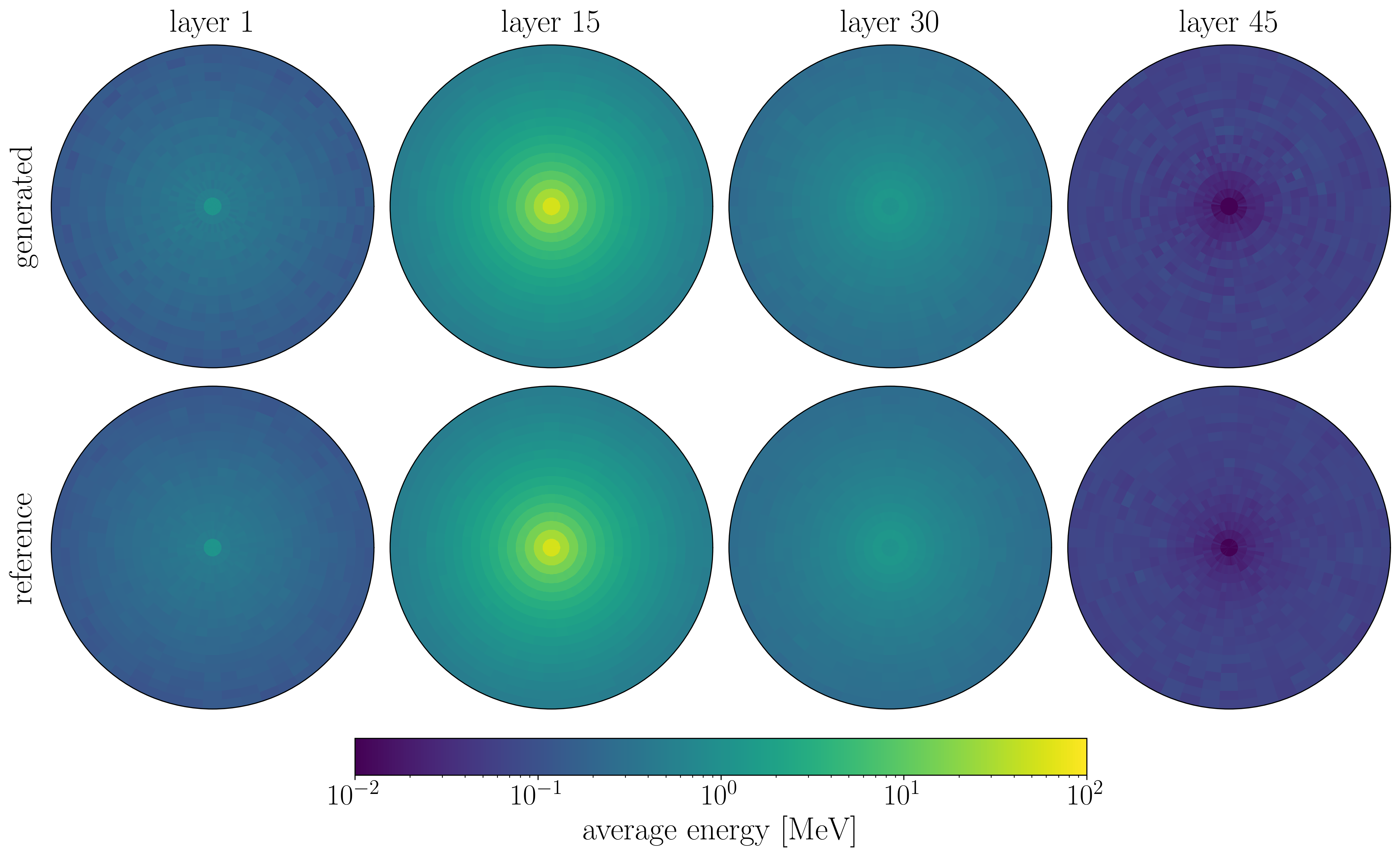}
    \caption{Average deposited energy for all cells in layer 1, 15, 30 and 45 overall flow generated (upper row) and \textsc{Geant4} simulated (lower row) sample. Shown are the results for \textsc{CaloChallenge} datast~3.}
    \label{fig:overlay_ds3}
\end{figure}

In the following, results are reported for the \textsc{GettingHigh} dataset and dataset~3 of the \textsc{CaloChallenge}. The full \(30\times30\times30\) voxel resolution of the \textsc{GettingHigh} dataset is used, which was not possible for the non-convolutional version of \textsc{L2LFlows}~\cite{Diefenbacher:2023vsw}. After training an instance of our model on both datasets, we draw 100,000 samples from both instances using the same incident energy distribution as the respective test data. These samples are then compared with the corresponding test data.

As a benchmark, the performance of \textsc{L2LFlows} is compared to the one of the \textsc{BIB-AE}~\cite{Buhmann:2020pmy}. The \textsc{BIB-AE} has an autoencoder-based architecture using additional adversarial critic networks and losses to improve fidelity. A kernel density estimation models the posterior distribution in latent space.

In real detectors, read-outs at small energies are dominated by noise.
Therefore, an energy cut-off of half the energy of a minimal ionizing particle is applied. Energy depositions below \(100\)~keV are removed before comparing simulated and generated showers on \textsc{GettingHigh}. To be consistent with the \textsc{CaloChallenge} evaluation, we use \(15.15\)~keV as the cut-off value on \textsc{CaloChallenge} dataset~3.

\subsection{Observables}
\label{sec:observables_r}
\begin{figure}
    \centering
    \includegraphics[width=0.35\linewidth]{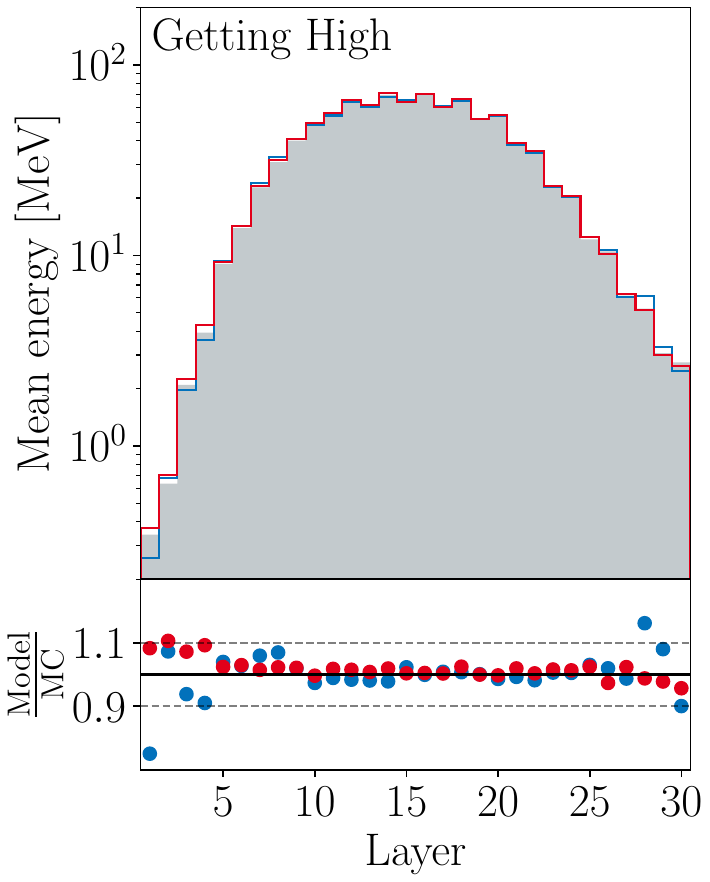}
    \includegraphics[width=0.35\linewidth]{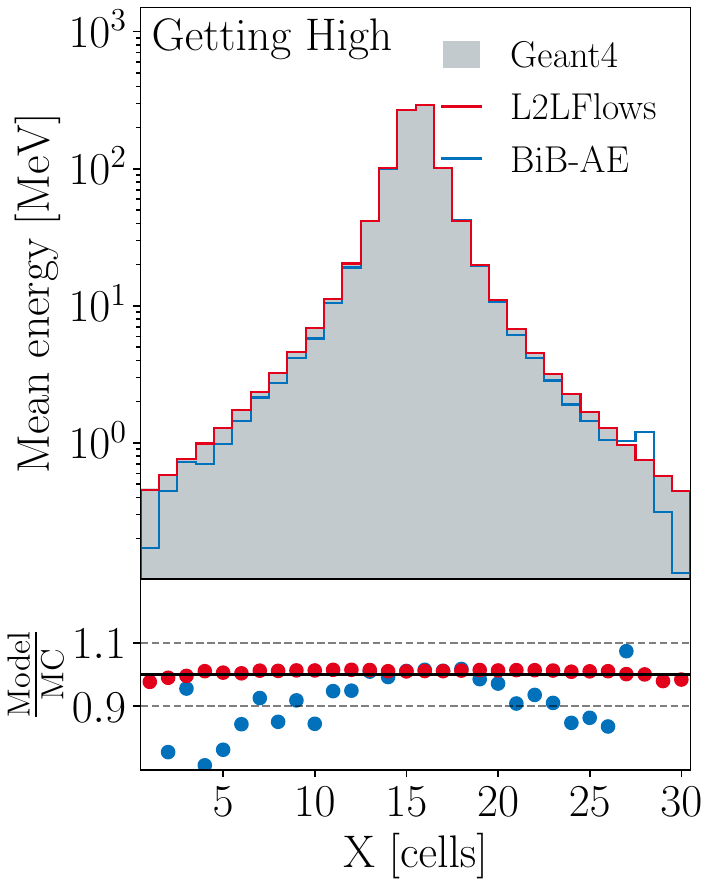}

    \includegraphics[width=0.35\linewidth]{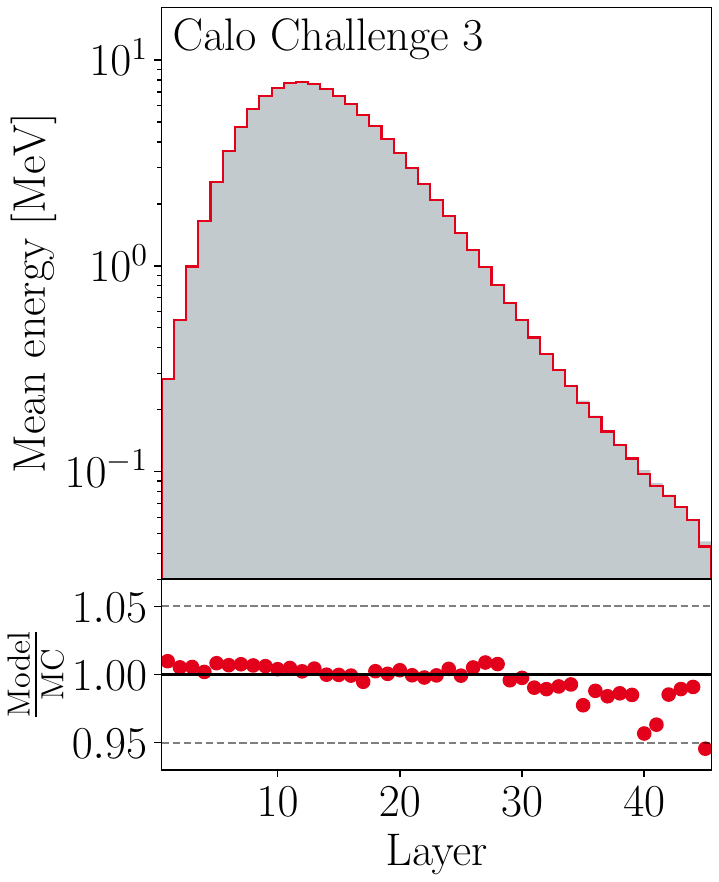}
    \includegraphics[width=0.35\linewidth]{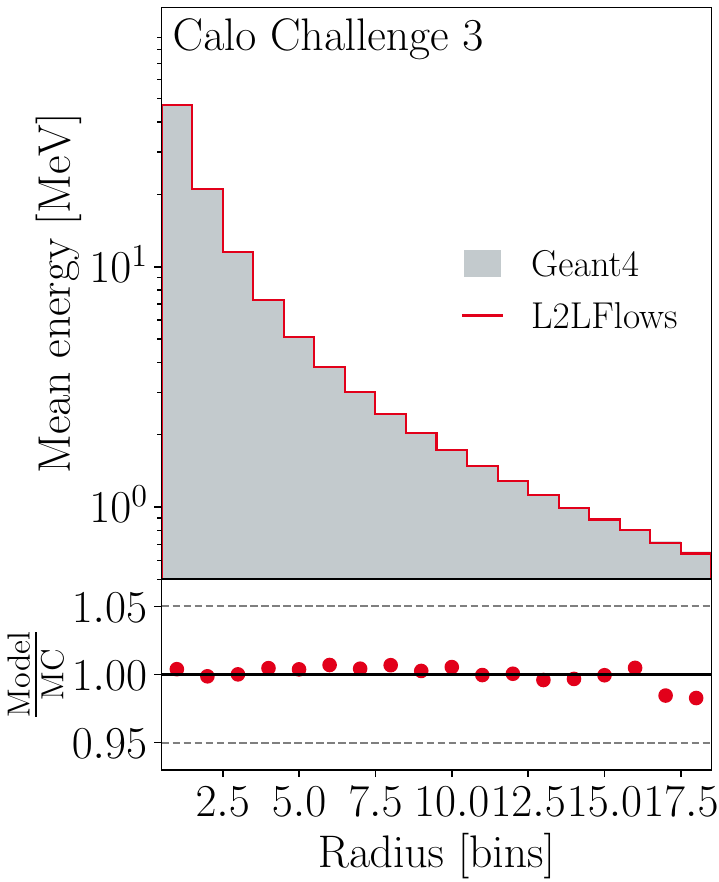}
    \caption{Top row: longitudinal (left) and \(X\) profile for \textsc{GettingHigh}. Bottom row: longitudinal (left) and radial (right) profile for \textsc{CaloChallenge} dataset~3.}
    \label{fig:profile}
\end{figure}
\begin{figure}
    \centering
    \includegraphics[width=0.7\linewidth]{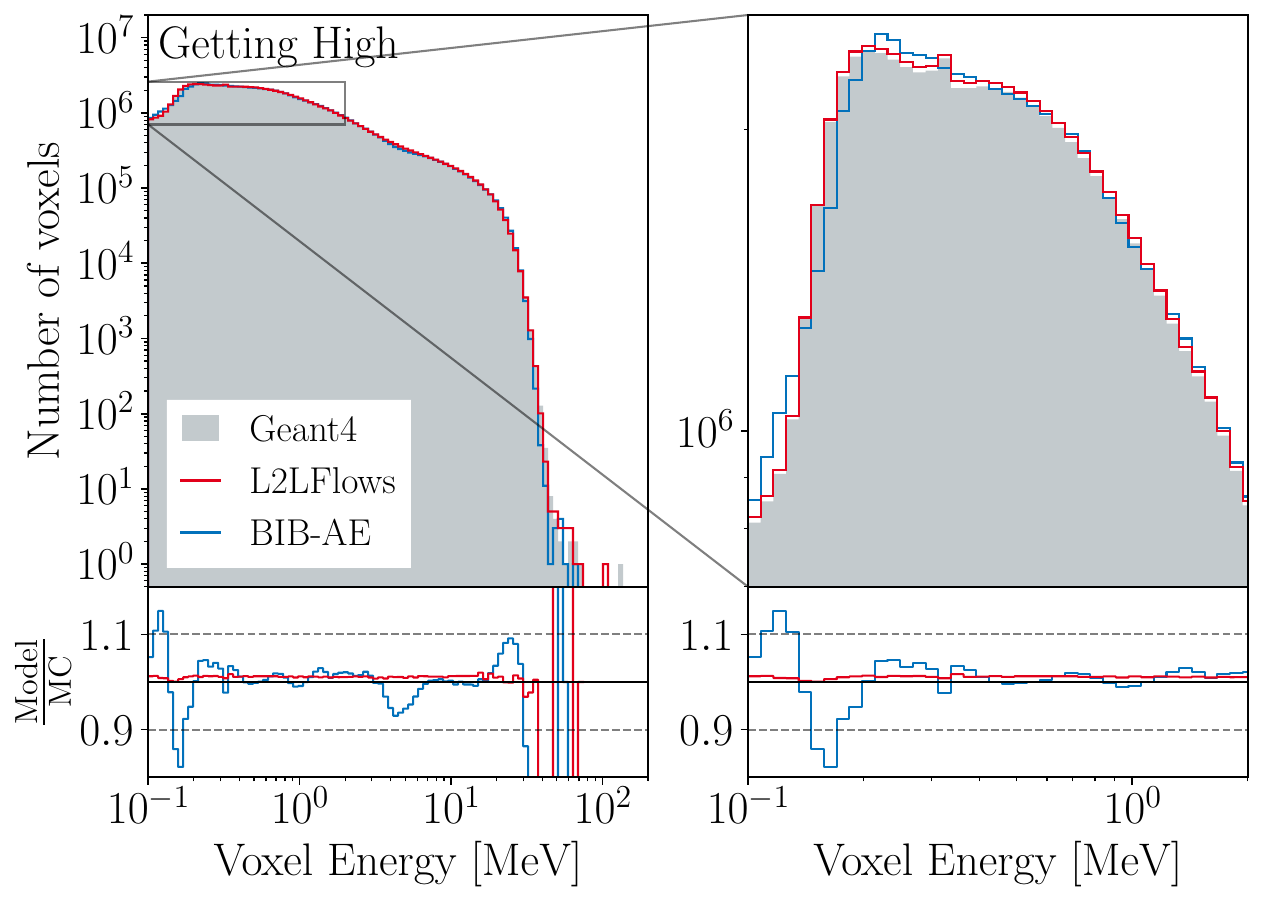}

    \includegraphics[width=0.7\linewidth]{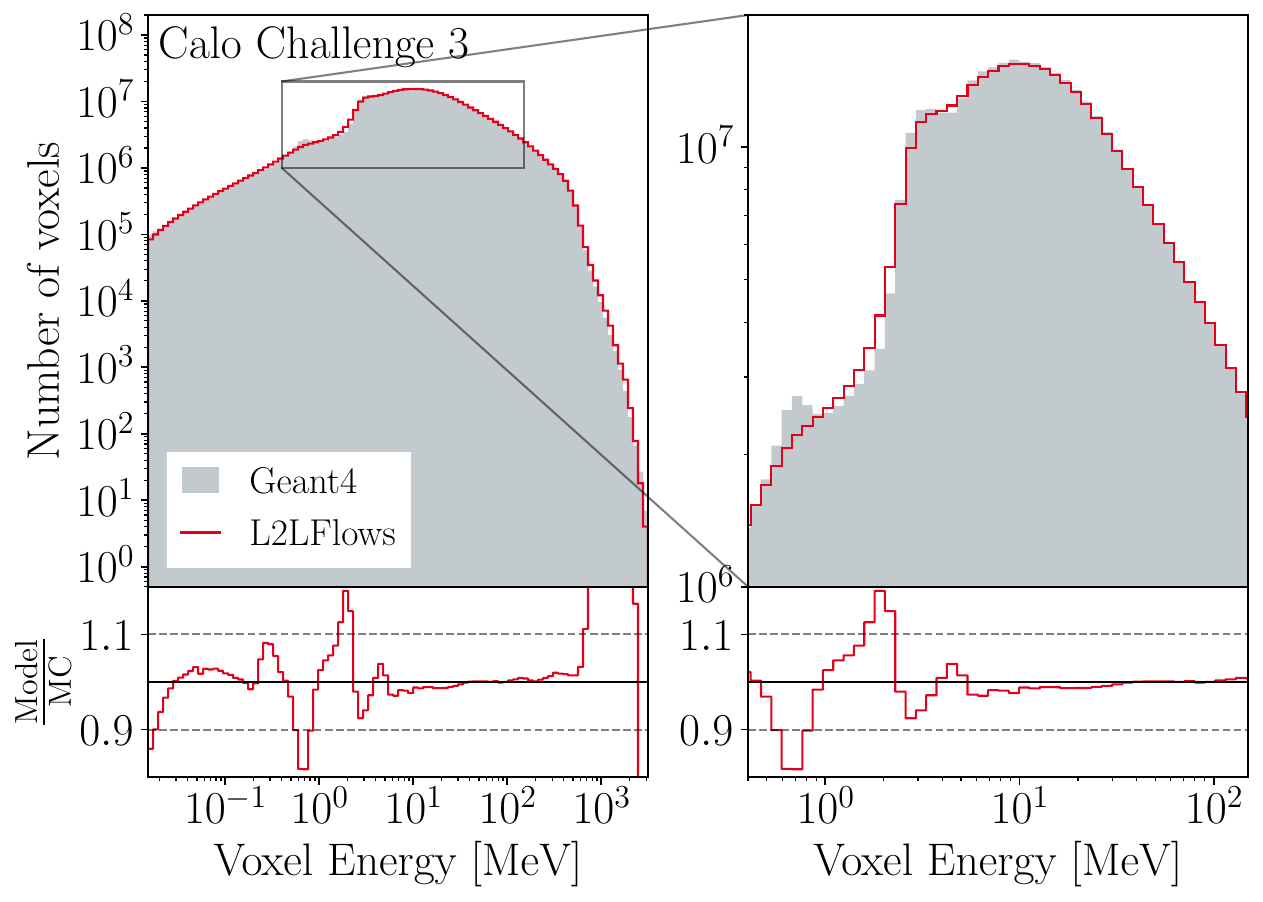}
    \caption{Cell energy spectrum.}
    \label{fig:spectrum}
\end{figure}
\begin{figure}
    \centering
    \includegraphics[width=0.35\linewidth]{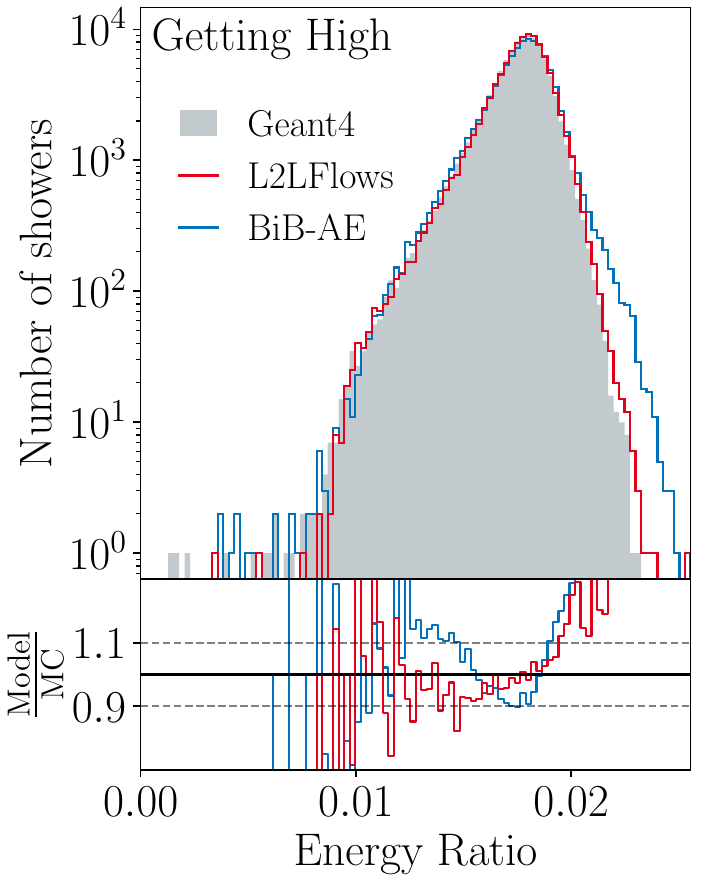}
    \includegraphics[width=0.35\linewidth]{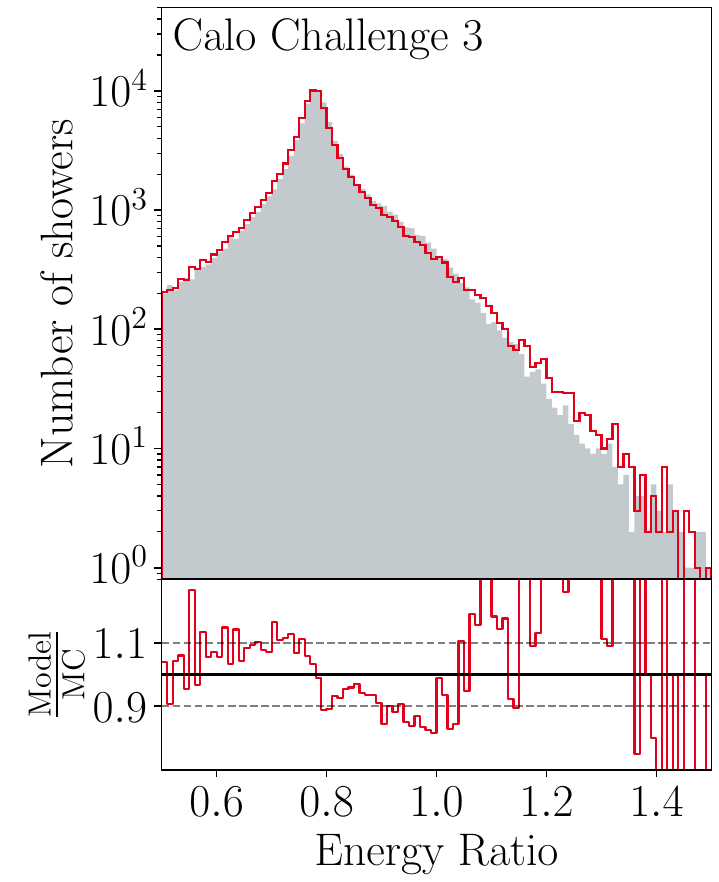}

    \includegraphics[width=0.35\linewidth]{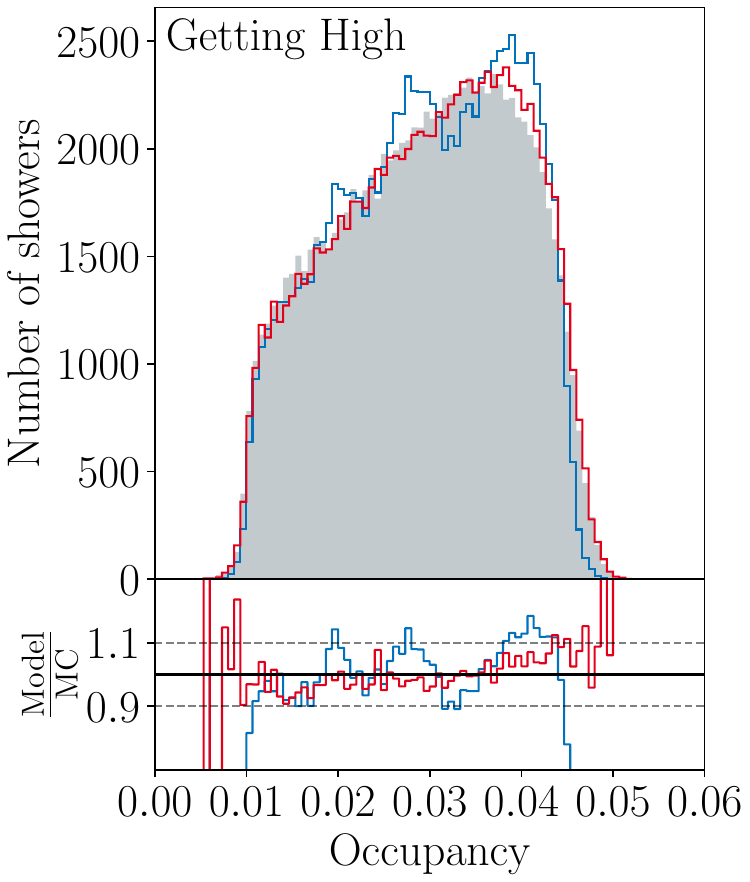}
    \includegraphics[width=0.35\linewidth]{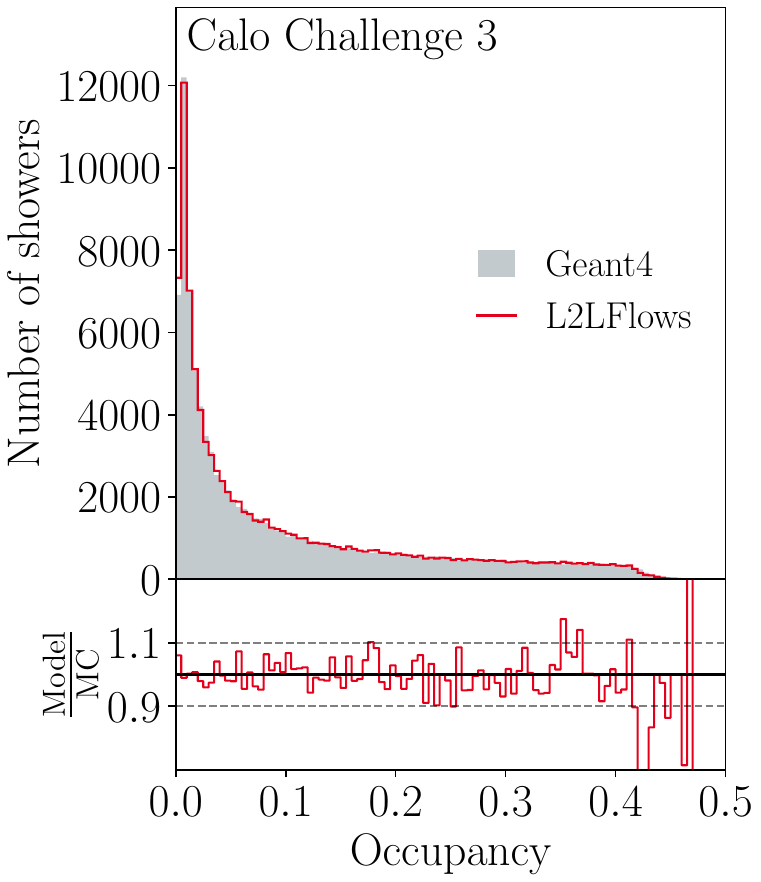}
    \caption{Top row: Total deposited energy over incident energy for \textsc{GettingHigh} (left) and \textsc{CaloChallenge} dataset~3 (right). The voxel energies in \textsc{CaloChallenge} dataset~3 are rescaled to account for the sampling fraction of the calorimeter. This explains the values above one and why the values for \textsc{GettingHigh} are way lower. Bottom row: Occupancy for \textsc{GettingHigh} (left) and \textsc{CaloChallenge} dataset~3 (right)}
    \label{fig:ratio}
\end{figure}
Figure~\ref{fig:profile} shows the longitudinal and \(X\) profile for \textsc{GettingHigh} and the longitudinal and radial profile for \textsc{CaloChallenge} dataset~3. We see an excellent agreement between generated and test samples, with deviations well below five percent. The only larger deviation can be found in the first four layers of \textsc{GettingHigh}. The \textsc{BIB-AE} also agrees well with the test data. However, it can not model the \(X\) profile as accurately as \textsc{L2LFlows} does.

Figure~\ref{fig:spectrum} shows the voxel energy spectrum. Here, we also see good agreements. In regions with sharp features in the reference distribution, deviations sometimes exceed \(10\%\). We get a better agreement for \textsc{GettingHigh} than for \textsc{CaloChallenge} dataset~3 because post-processing is applied only for \textsc{GettingHigh}.

Further especially important observables are the energy ratio and the occupancy. Histograms for these observables are shown in figure~\ref{fig:ratio}. We see good agreements in the bulge of the distributions. In the tail regions, deviations are larger. We see an improvement compared to the \textsc{BIB-AE} performance. Further histograms can be found in~\ref{sec:histograms}.

\begin{figure}
    \centering
    \includegraphics[width=0.48\linewidth]{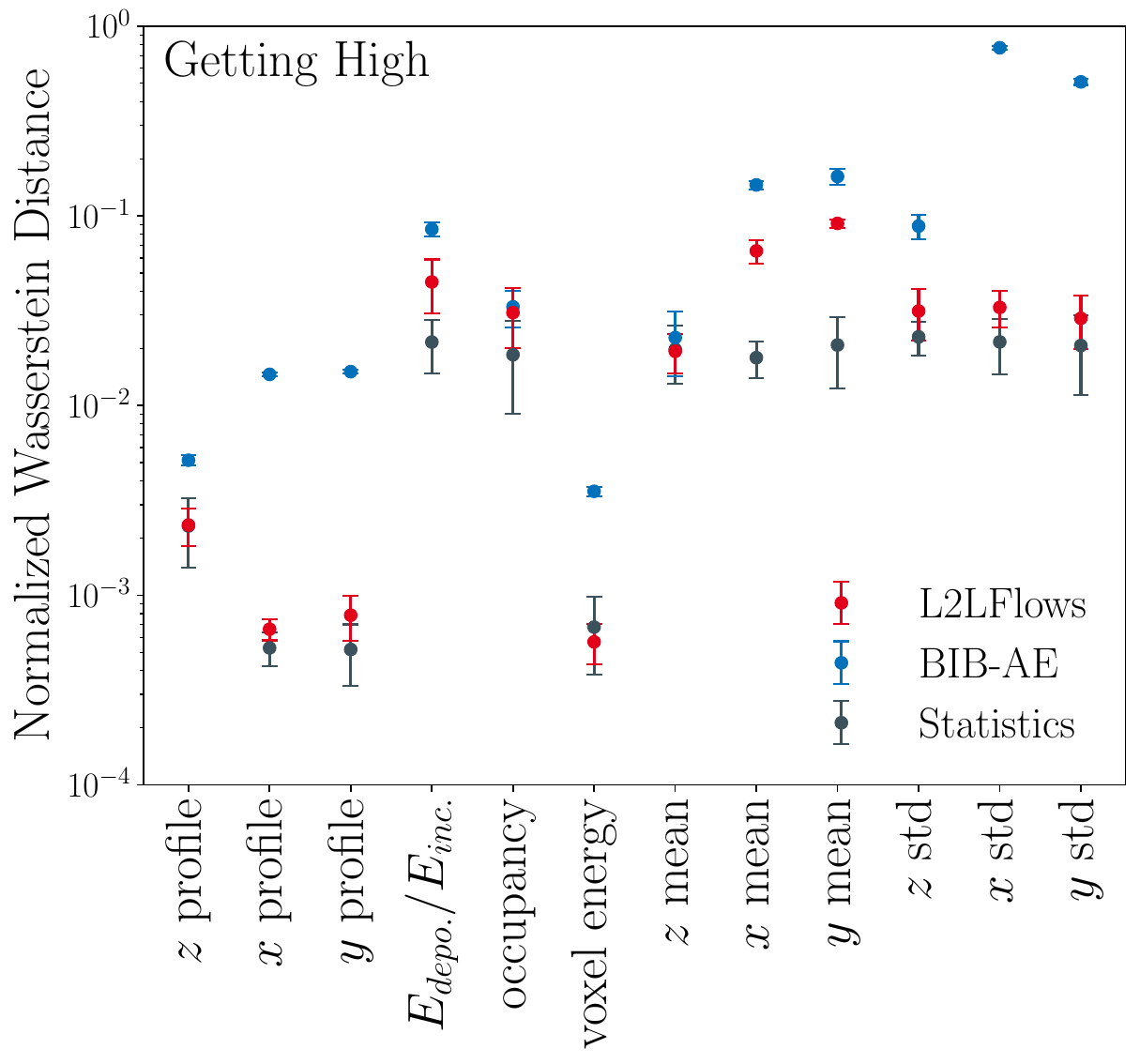}
    \includegraphics[width=0.48\linewidth]{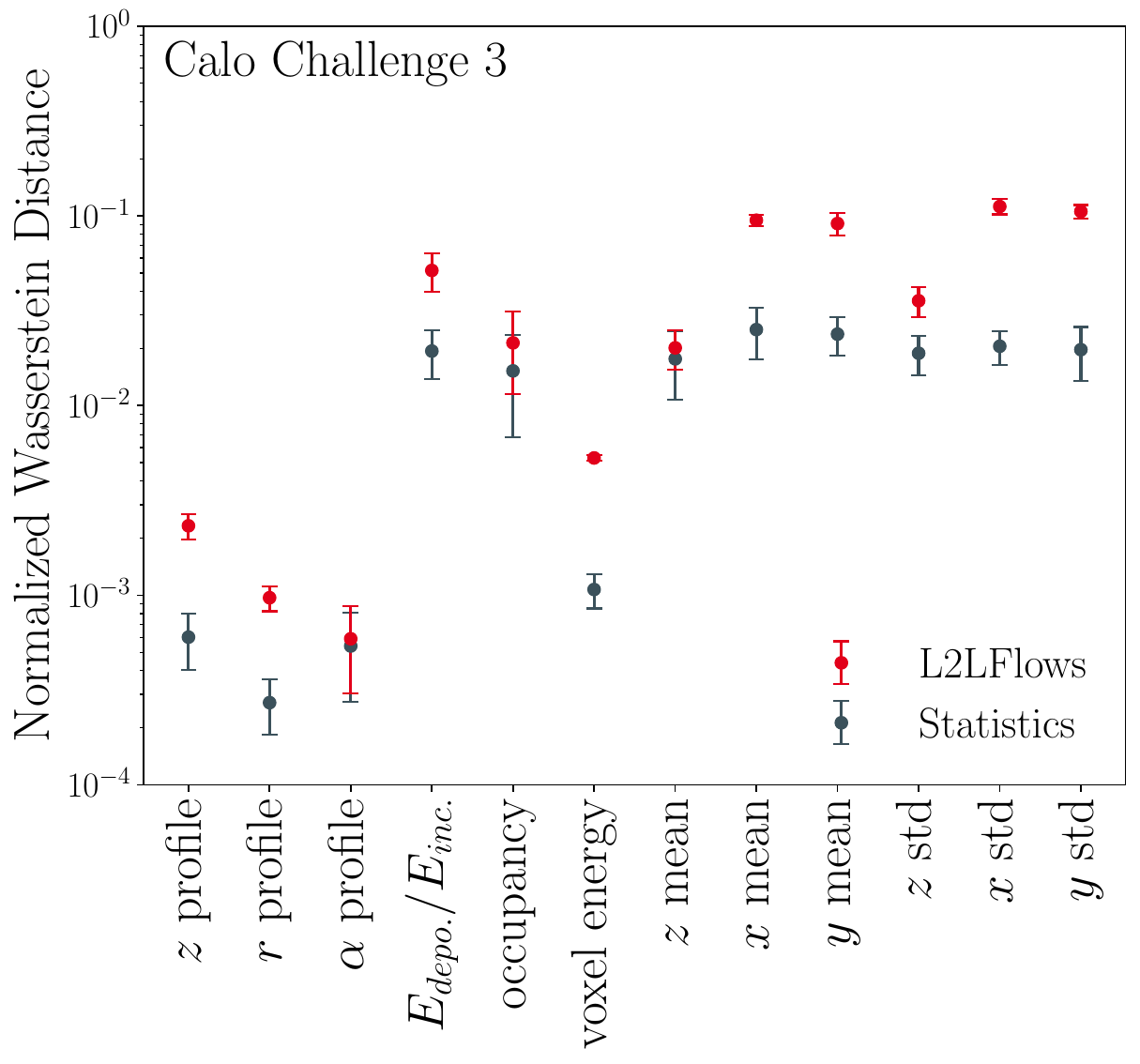}
    \caption{Wasserstein distances for several physics observables between generated and test data. The Wasserstein distances are numerically evaluated using ten batches consisting of 10,000 generated and 10,000 test set showers. Shown are the mean and the standard deviation over the ten resulting values. The gray markers show \textsc{Geant4} train samples compared to \textsc{Geant4} test samples. It gives us the lowest Wasserstein distance that can be resolved numerically given the amount of statistics.}
    \label{fig:wasserstein}
\end{figure}
The diagrams in Figure~\ref{fig:wasserstein} show numerical evaluated one-dimensional Wasserstein distances between test data and generated data. Given are the Wasserstein distances for all observables described in section~\ref{sec:observables_m}. To make the Wasserstein distance dimensionless, the distributions are standard scaled using the mean and standard deviation of the \textsc{Geant4} distribution. We see a clear improvement compared to the \textsc{BIB-AE} in almost all observables. We also notice that for many observables, there are no differences between \textsc{Geant4} and Flow-generated samples that can be resolved with the given amount of statistics.

\subsection{Classifier}
\label{sec:classifier}
\begin{table}[t]
    \centering
    \begin{tabular}{ l l r r r r}
        \toprule
        &&\multicolumn{2}{r}{high level classifier}&\multicolumn{2}{r}{low level classifier}\\
        Dataset & Simulator & AUC & JSD & AUC & JSD \\
        \midrule
         \textsc{GettingHigh} & \textsc{L2LFlows} & \(.634\pm.002\) & \(.047\pm.002\) & \(.905\pm.003\) & \(.438\pm.009\) \\
         & \textsc{BIB-AE} & \(.903\pm.002\) & \(.436\pm.005\) & \(\gg .999\) & \(.985\pm.001\) \\
        \midrule
        \textsc{CaloChallenge~3} & \textsc{L2LFlows} & \(.686\pm.002\) & \(.084\pm.001\) & \(.983\pm.002\) & \(.760\pm.013\)\\
        \bottomrule
    \end{tabular}
    \caption{Results of the classifier test. Shown are the mean and standard deviation over five random network initializations.}
    \label{tab:classifier}
\end{table}
We also conducted a classifier test to investigate multidimensional, potentially non-linear correlations.  See details on this in section~\ref{sec:classifier_m}

The results of our classifier study are summarized in table~\ref{tab:classifier}. In consistency with the  Wasserstein metric, we see a clear improvement of our flow model compared to the \textsc{BIB-AE}. The JSD value of 0.047 for our flow model on \textsc{GettingHigh} indicates an excellent agreement of the distributions in the nine-dimensional input space of our high-level classifier. The low-level classifier can distinguish \textsc{BIB-AE} and \textsc{Geant4} samples with almost perfect accuracy. AUC and JSD values are significantly lower for the flow-generated samples, indicating an overlap of the distributions.

\subsection{Timing}
\label{sec:timing}
\begin{table}
    \centering
    \begin{tabular}{ l l r r r}
        \toprule
        Simulator & Hardware & Batch size & time [ms] & Speedup \\
        \midrule
        \textsc{Geant4} & CPU & 1 & 4081.53 \(\pm\) \makebox[1.2cm][r]{169.92} & x1.0 \\
        \midrule
        \textsc{L2LFlows} & CPU & 1 & 1202.66 \(\pm\) \makebox[1.2cm][r]{26.11} & x3 \\
        & & 10 & 514.49 \(\pm\) \makebox[1.2cm][r]{0.33} & x8 \\
        & & 100 & 417.55 \(\pm\) \makebox[1.2cm][r]{6.73} & x10 \\
        & & 1000 & 488.75 \(\pm\) \makebox[1.2cm][r]{3.47} & x8 \\
        \midrule
        \textsc{BIB-AE} & CPU & 1 & 426.32 \(\pm\) \makebox[1.2cm][r]{3.62} & x10 \\
        & & 10 & 424.71 \(\pm\) \makebox[1.2cm][r]{3.53} & x10 \\
        & & 100 & 418.04 \(\pm\) \makebox[1.2cm][r]{0.20} & x10 \\
        \midrule
        \textsc{L2LFlows} & GPU & 1 & 1155.01 \(\pm\) \makebox[1.2cm][r]{23.44} & x4 \\
        & & 10 & 118.10 \(\pm\) \makebox[1.2cm][r]{2.24}  & x35 \\
        & & 100 & 12.34 \(\pm\) \makebox[1.2cm][r]{0.58} & x331 \\
        & & 1000 & 3.24 \(\pm\) \makebox[1.2cm][r]{0.05} & x1260 \\
        \midrule
        \textsc{BIB-AE} & GPU & 1 & 3.19 \(\pm\) \makebox[1.2cm][r]{0.01} & x1279 \\
        & & 10 & 1.57 \(\pm\) \makebox[1.2cm][r]{0.01} & x2600  \\
        & & 100 & 1.42 \(\pm\) \makebox[1.2cm][r]{0.01} & x2874 \\
        \bottomrule
    \end{tabular}
    \caption{Timing on \textsc{GettingHigh} with incident energies uniformly sampled between \(10\) and \(100\)~GeV. Shown are the times for our flow-based model, for the \textsc{BIB-AE} and for \textsc{Geant4}. The CPU version was run on a single thread of a Intel\textsuperscript{\textregistered} Xeon\textsuperscript{\textregistered} CPU E5-2640 v4 and the GPU version on a NVIDIA\textsuperscript{\textregistered} V100. Given are mean and standard deviation over 25 runs. The \textsc{BIB-AE} and \textsc{Geant4} times are taken from \cite{Buhmann:2020pmy}.}
    \label{tab:timing}
\end{table}
The main goal of training generative networks on calorimeter showers is to gain a speedup over Monte-Carlo-based simulators. To test how much faster our model can generate showers than \textsc{Geant4} does, we perform a timing study on \textsc{GettingHigh}. To be able to compare our times with the findings of the authors of the \textsc{BIB-AE}, the same hardware was used. The times for \textsc{Geant4} and the \textsc{BIB-AE} are taken from \cite{Buhmann:2020pmy}. To ensure comparability with \textsc{Geant4}, only a single execution thread is used for the CPU case.

If one compares our findings with the ones in \cite{Diefenbacher:2023vsw}, one notices that our flow is faster on 30x30x30 voxel showers than the original \textsc{L2LFlows} on 10x10x30 voxel shower cores. The reason is that the original \textsc{L2LFlows} is built using masked autoregressive blocks. For a data dimension of \(n\), these blocks are \(n\) times slower during generation than during training. We used coupling blocks instead, which can be executed forward and backward with the same speed.

Compared with the \textsc{BIB-AE}, for a batch size of 100, both models executed on a CPU are equally fast. For smaller batch sizes, the \textsc{BIB-AE} is faster. The reason is that with smaller batch sizes, vectorizing computations can be done less efficiently. The \textsc{L2LFlows} has to loop over all layers, which effectively reduces the batch size by a factor of the number of layers compared to models that do not have to loop over layers. This could be potentially overcome by compiling the model.

\section{Conclusion}
\label{sec:conclusion}
Based on \textsc{L2LFlows}~\cite{Diefenbacher:2023vsw} and \textsc{GLOW}~\cite{Kingma:2018gyr}, we have developed an enhanced flow-based model for calorimeter showers. Our approach significantly advances the scalability and fidelity of \textsc{L2LFlows} by switching from masked autoregressive flows to coupling flows with convolutional layers and \textsc{U-Net}-type connections in the sub-networks. This architecture addresses the challenge of high-dimensional data spaces that have traditionally limited the applicability of normalizing flow models in particle detector simulations.

Using convolutional layers reduces the number of parameters and leads to better scaling with cardinality. The \textsc{U-Net} architecture allows the model to learn features on different scales. Furthermore, by transitioning from masked autoregressive flows to coupling flows, we achieve faster inference times without sacrificing the quality of the generated samples~\cite{Ernst:2023qvn}.

Empirical validations using the \textsc{GettingHigh} dataset demonstrate that --- in line with the results of~\cite{Diefenbacher:2023vsw} --- our model reaches and often exceeds the fidelity of state-of-the-art models like the \textsc{BIB-AE}. Wasserstein distances between generated and test data are significantly lower for our model than for the \textsc{BIB-AE}. A classifier-based study confirms this result. Timing the model on a CPU and a GPU shows that our model can generate calorimeter showers faster than traditional simulations. The model is as fast as the \textsc{BIB-AE} on CPU with a batch size of 100. For smaller batch sizes, it gets slower.

Looking ahead, our research opens up several avenues for further study. One could investigate ways to speed up the model further, for instance, by transforming it into a latent model. Another crucial direction would be to integrate the model into a full simulation chain and assess its impact on physics analyses. Also, the model could be applied to hadronic showers, which are more complex than electromagnetic showers. \textsc{Convolutional L2LFlows} was submitted to the Fast Calorimeter Simulation Challenge for datasets~2 and 3 where a detailed comparison between recent models will be done.

In conclusion, our work contributes to the evolution of generative modeling in high-energy physics, presenting a robust, scalable, and accurate method for modeling particle showers in complex detectors. As the field progresses towards more granular calorimeters and higher event rates, the importance of such advanced computational methods will only increase.

\section*{Code Availability}
The code for this study can be found under \url{https://github.com/FLC-QU-hep/ConvL2LFlow}.

\section*{Acknowledgments}
We thank Sascha Diefenbacher for early contributions to this work. We want to thank Henry Day-Hall for valuable comments on the manuscript.

This research was supported in part by the Maxwell computational resources operated at Deutsches Elektronen-Synchrotron DESY, Hamburg, Germany. This project has received funding from the European Union’s Horizon 2020 Research and Innovation programme under Grant Agreement No 101004761. We acknowledge support by the Deutsche Forschungsgemeinschaft under Germany’s Excellence Strategy – EXC 2121  Quantum Universe – 390833306 and via the KISS consortium (05D23GU4, 13D22CH5) funded by the German Federal Ministry of Education and Research BMBF in the ErUM-Data action plan. CK would like to thank the Baden-W\"urttemberg-Stiftung for financing through the program \textsl{Internationale Spitzenforschung}, pro\-ject \textsl{Uncertainties – Teaching AI its Limits} (BWST\_IF2020-010). DS is supported by the U.S. Department of Energy (DOE), Office of Science grant DOE-SC0010008. 

\bibliographystyle{JHEP}
\bibliography{literature}

\newpage
\appendix
\section{Architecture and Training}
\label{sec:architecture}
We here only give hyper-parameter values since we describe the architecture and training in the main text. These parameters are given in \cref{tab:hp_energy_flow,tab:hp_energy_flow_training,tab:hp_causal_flows,tab:hp_u_net,tab:hp_embedding,tab:hp_causal_flows_training}. We did some hyper-parameter optimization by hand, training a single of the causal flows. For automated optimization, the training is too resource-intensive.

In table~\ref{tab:num_parameters}, we give the total number of trainable parameters and the number of parameters involved in the generation process.
\begin{table}[H]
    \centering
    \begin{tabular}{l l}
        \toprule
        parameter & value \\
        \midrule
        block type & MADE \\
        permutation & random \\
        \#blocks & 6 \\
        \#hidden layers & 1 \\
        \#hidden features & 64 \\
        spline type & rational quadratic \\
        spline bins & 8 \\
        minimum bin height & \(10^{-6}\) \\
        minimum bin width & \(10^{-6}\) \\
        minimum derivative & \(10^{-6}\) \\
        transformation interval & \([\text{logit}(10^{-6}), \text{logit}(1-10^{-6})]\) \\
        base distribution & Gaussian \\
        \bottomrule
    \end{tabular}
    \caption{Hyper-parameters of the energy distribution flow architecture}
    \label{tab:hp_energy_flow}
\end{table}
\begin{table}[H]
    \centering
    \begin{tabular}{l l}
        \toprule
        parameter & value \\
        \midrule
        \#epochs (\textsc{CaloChallenge~3}) & 2000 \\
        \#epochs (\textsc{GettingHigh}) & 200 \\
        learning rate scheduler & exponential \\
        initial learning rate & \(10^{-3}\) \\
        final learning rate & \(10^{-6}\) \\
        batch size & 1024 \\
        additional noise distribution & Gaussian \\
        additional noise mean & 1~keV \\
        additional noise std & 0.2~keV \\
        \bottomrule
    \end{tabular}
    \caption{Hyper-parameters of the energy distribution flow training}
    \label{tab:hp_energy_flow_training}
\end{table}
\begin{table}[H]
    \centering
    \begin{tabular}{l l}
        \toprule
        parameter & value \\
        \midrule
        squeezing factor & 2 \\
        block type & coupling \\
        permutation & \(1\times{}1\) convolution \\
        normalization & activation norm \\
        \#blocks & 8 \\
        spline type & rational quadratic \\
        spline bins & 8 \\
        minimum bin height & \(10^{-6}\) \\
        minimum bin width & \(10^{-6}\) \\
        minimum derivative & \(10^{-6}\) \\
        transformation interval & \([-15,15]\) \\
        base distribution & Gaussian \\
        sub network & U-Net \\
        embedding network & convolutional \\
        \bottomrule
    \end{tabular}
    \caption{Hyper-parameters of the causal flows architecture}
    \label{tab:hp_causal_flows_training}
\end{table}
\begin{table}[H]
    \centering
    \begin{tabular}{l l l}
        \toprule
        parameter & value \textsc{CaloChallenge~3} & value \textsc{GettingHigh} \\
        \midrule
        \#feature maps in hidden layers & 32 & 32 \\
        first down sampling kernel size & (3,5) & (3,3) \\
        second down sampling kernel size & (3,5) & (5,5) \\
        use cyclic padding & true & false \\
        activation & leaky ReLu & leaky ReLu \\
        \bottomrule
    \end{tabular}
    \caption{U-Net hyper parameters}
    \label{tab:hp_u_net}
\end{table}
\begin{table}[H]
    \centering
    \begin{tabular}{l l l l l}
        \toprule
        layer & out channels & kernel size & stride & padding \\
        \midrule
        convolution 2d & 8 & 1 & 1 & 0 \\
        convolution 2d & 8 & 3 & 1 & 1 \\
        convolution 2d & 8 & 2 & 2 & 0 \\
        convolution 2d & 8 & 3 & 1 & 1 \\
        \bottomrule
    \end{tabular}
    \caption{Embedding network layer. This architecture is used for all but the first causal flow. For \textsc{CaloChallenge~3} the padding is cyclic in the last dimension. As activaion a leaky ReLu with a negative slope of \(0.01\) is used.}
    \label{tab:hp_embedding}
\end{table}
\begin{table}[H]
    \centering
    \begin{tabular}{l l l}
        \toprule
        parameter & value \textsc{CaloChallenge~3} & value \textsc{GettingHigh} \\
        \midrule
        \#epochs & 800 & 200 \\
        learning rate scheduler & one cycle & one cycle \\
        maximum learning rate & \(6\times10^{-4}\) & \(10^{-4}\) \\
        batch size & 1024 & 1024 \\
        additional noise distribution & uniform & uniform \\
        additional noise width & 0.1~keV & 0.1~keV \\
        zero pixel fill distribution & log normal & log normal \\
        zero pixel fill median & 1~keV & 1~keV \\
        zero pixel fill std (in \(\log_{10}\) space) & 0.2 & 0.2 \\
        \(L2\) gradient clipping & 100 & 800 \\
        \(L2\) weight decay & 0 & 0.1 \\
        \bottomrule
    \end{tabular}
    \caption{Hyper-parameters of the causal flows training}
    \label{tab:hp_causal_flows}
\end{table}
\begin{table}[H]
    \centering
    \begin{tabular}{l r r}
        \toprule
        model & total & generation \\
        \midrule
        \textsc{L2LFlows} & 163,269,320 & 163,269,320 \\
        \textsc{BIB-AE} & 76,515,284 & 36,517,632 \\
        \textsc{L2LFlows} (\textsc{CaloChallenge~3}) & 194,964,482 & 194,964,482 \\
        \bottomrule
    \end{tabular}
    \caption{Number of parameters of the models. middle: total number of parameters, right: number of parameters used for generation.}
    \label{tab:num_parameters}
\end{table}

\section{Preprocessing}
\label{sec:preprocessing}
The inputs for our network are energies that range over several orders of magnitude. This makes it hard for neural networks to learn features. An appropriate rescaling of the input helps the network. We use a combination of log, logit, and affine transformations. The log transformation helps the network to deal with inputs distributed over different orders of magnitude, while the logit transformation helps the flow to only generate values in the correct range. Since all three transformations are easily invertible during generation, we can map the generated samples back to physics space.

As in previous works we used a modified logit transformation given by
\begin{align}
    \text{logit}_{\alpha}(x) &= \log\left(\frac{x\times(1-2\alpha)+\alpha}{1-x\times(1-2\alpha)-\alpha}\right)
\end{align}
where \(\alpha\) is a free parameter. For the energy distribution flow the layer energies are preprocessed in the following way
\begin{align}
    E_{\text{max}} &= \max_{\text{training data}}\max_{i} E_{i} \\
    x_{i} &= \text{logit}_{10^{-6}}\left(\frac{E_i}{E_{\text{max}}}\right)
\end{align}
where \(E_{i}\) is the energy deposited in layer \(i\) and \(x\) is the input for the flow. The causal flows input is also preprocessed
\begin{align}
    I_{\text{max},i} &= \max_{\text{training data}}\max_{jk} I_{ijk} \\
    x_{ijk} &= \text{logit}_{10^{-2}}\left(1+\frac{\log\left(10^{-7}+\frac{I_{ijk}}{v_{\text{max},i}}\right)}{\log(10^{7})}\right)
\end{align}
where \(I_{ijk}\) denote the voxel in layer \(i\) with the position \(j,k\) and \(x\) is the input for flows. The conditional inputs are also preprocessed
\begin{align}
    \text{condition}_{inc} &= \log\left(10^{-6}+\frac{E_{inc}}{1~\text{GeV}}\right) \\
    \text{condition}_{i} &= \log\left(10^{-6}+\frac{E_{i}}{1~\text{GeV}}\right)
\end{align}
where \(E_{inc}\) is the incident energy and \(E_i\) is the energy deposited in layer \(i\).

\section{Postprocessing}
\label{sec:postprocessing}
This appendix describes how we map the generated voxel energy spectrum onto the simulated one. The voxel energy spectrum can be modified by applying voxel-wise transformations to the shower. We construct this function and numerically fit a spline to it. Since this function is close to the identity, we do not expect a large effect on other observables.

We aim to map the generated voxel energy spectrum density \(p\) onto the \textsc{Geant4} one \(q\). Since we are only interested in active voxels, we do not consider voxel energies below the threshold. We evaluate everything in log space to make our distributions more uniform and avoid numerical instabilities.

The function \(f\) we want to construct is supposed to map \(p\) onto \(q\). According to the change of variable formula, this is equivalent to
\begin{equation}
	p(x) = q(f(x))f'(x).
\end{equation}
Two continuous functions fulfill this requirement: one monotonically increasing and one monotonically decreasing. We are going to construct the increasing function. If we integrate over the change of variable formula, we get
\begin{align}
	P(x) &= Q(f(x)) \\
	\Leftrightarrow f(x) &= Q^{-1}(P(x))
\end{align}
where \(P\) and \(Q\) are the cumulative density functions of \(p\) and \(q\).

Cumulative density functions evaluated at position \(x\) return by definition the fraction of points lower than \(x\). We randomly draw the same amount of points from both distributions and store them in two lists. By sorting both lists, we get a look-up table for \(P^{-1}\) and \(Q^{-1}\), and by matching the two lists, we get points in \(\mathbb{R}^2\) following the graph of \(f\).

To be able to evaluate \(f\) fast and not overfit to statistical noise, we split the interval into 128 bins of the same size and evaluate the mean of \(x\) and \(f(x)\) in every bin than we fitted a Cubic spline \(c\)  through these mean points. Since we evaluated everything in log space, the function we need to apply to the generated voxel energies is \(\exp \circ c \circ \log\). After fitting the spline, we discarded the generated showers and draw newly sampled showers from our model for evaluation.

\section{Classifier}
\label{sec:classifier_a}
Our high-level classifier has four hidden layers with 512, 512, 128, and 32 nodes and one output node. We use 100,000 generated samples per generator and 100,000 \textsc{Geant4} samples, which are not used to train the generators. A standard scaling of the inputs is applied. After an 80/20 split, we have 160,000 train and 40,000 test samples. We train the classifier for 30 epochs using binary cross entropy as loss function and ADAM as optimizer.

The architecture of our low-level classifier consists of four convolutional and four linear layers. We use maxpooling as the downsampling operation. A 60/20/20 split gives us 120,000 training, 40,000 validation, and 40,000 test samples. As for the high-level case, we train the classifier for 30 epochs using binary cross entropy as loss function and ADAM as optimizer. After training, we take the epoch with the lowest validation loss and fit a calibration curve~\cite{Niculescu:2005} using the validation set.

\section{Further Histograms}
\label{sec:histograms}
\begin{figure}[h]
    \centering
    \includegraphics[width=0.3\linewidth]{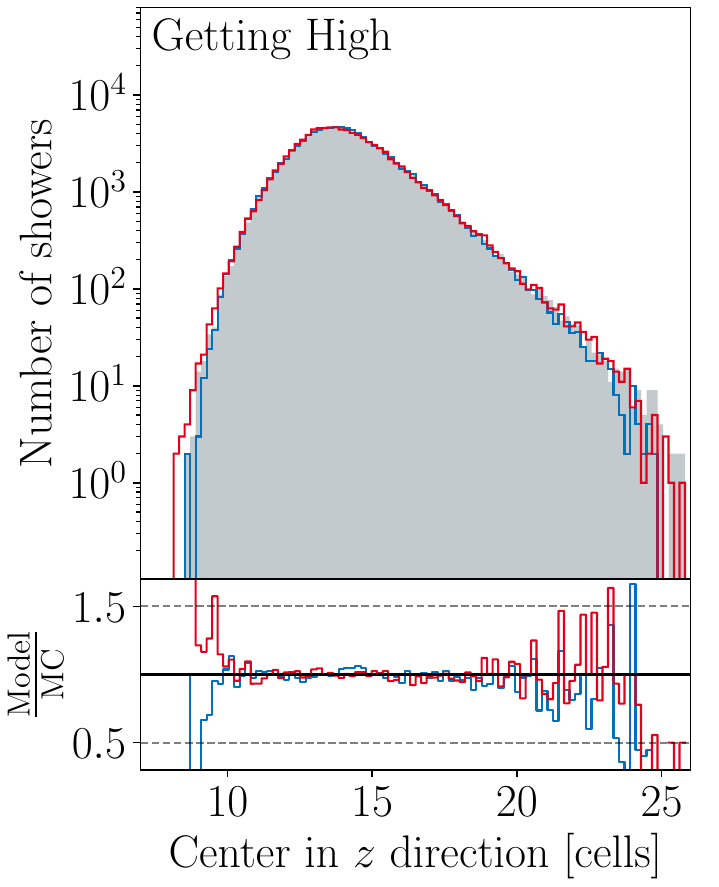}
    \includegraphics[width=0.3\linewidth]{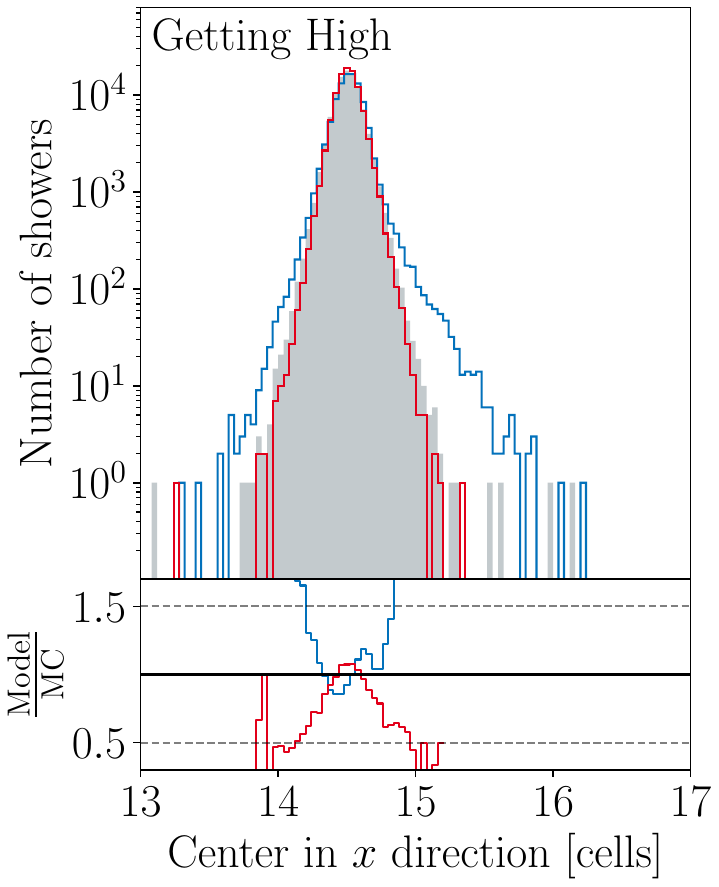}
    \includegraphics[width=0.3\linewidth]{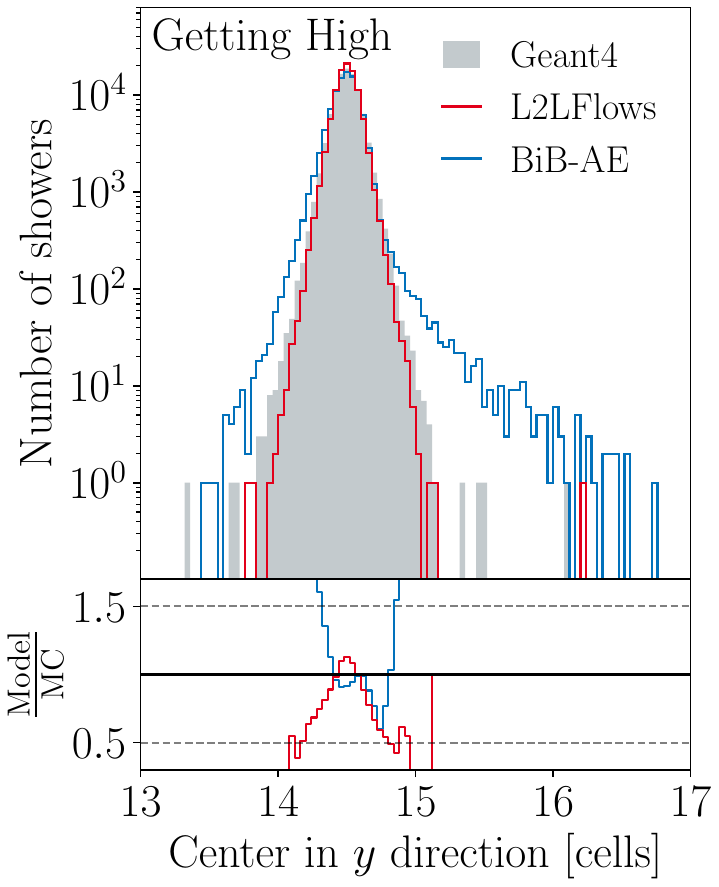}
    \caption{Center of gravity in longitudinal (left), \(x\) (middle), and \(y\) (left) direction for \textsc{GettingHigh}.}
\end{figure}
\begin{figure}[h]
    \centering
    \includegraphics[width=0.3\linewidth]{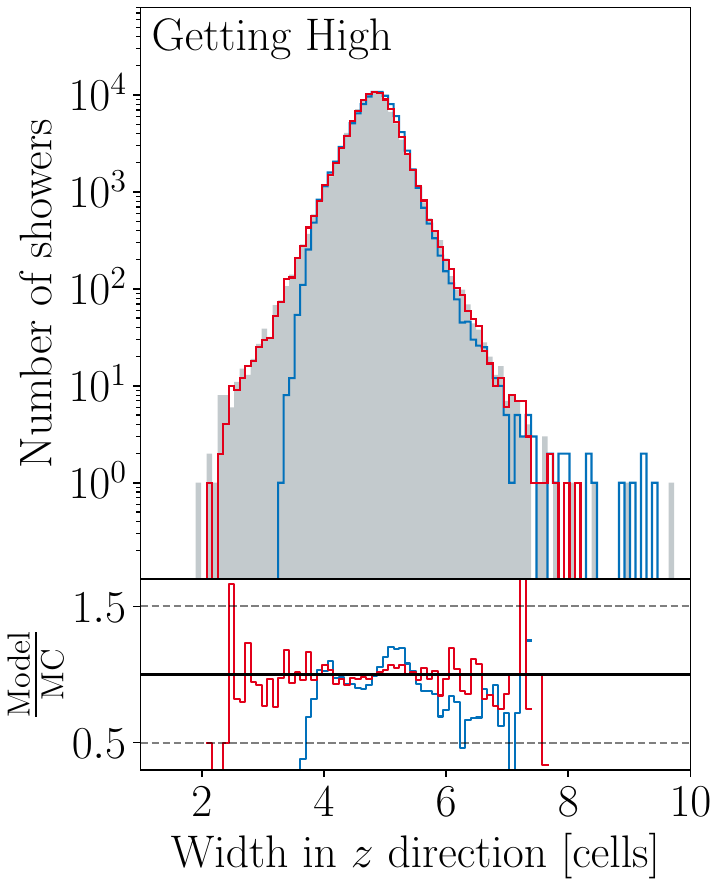}
    \includegraphics[width=0.3\linewidth]{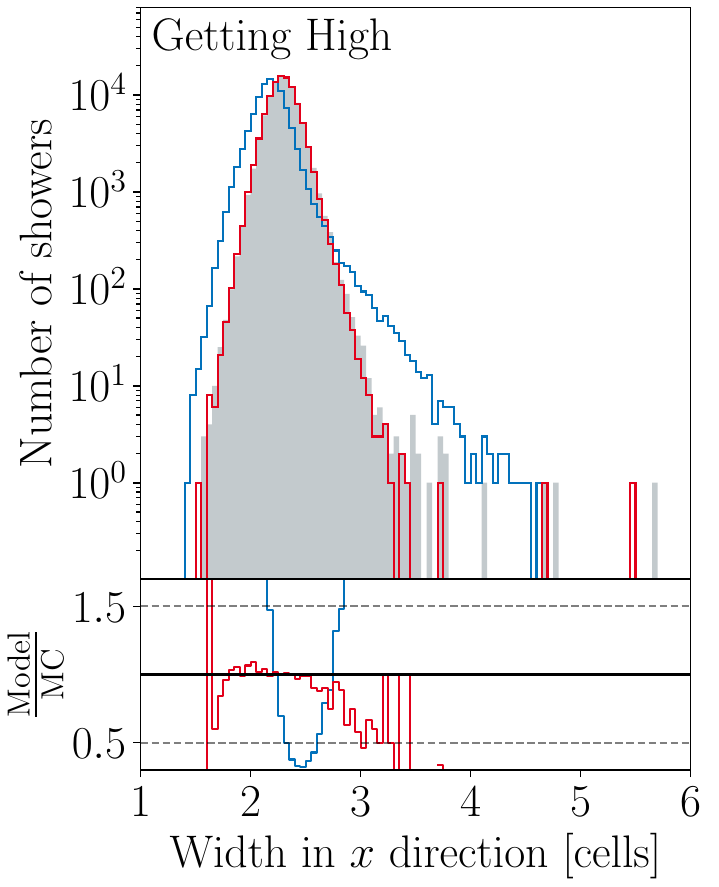}
    \includegraphics[width=0.3\linewidth]{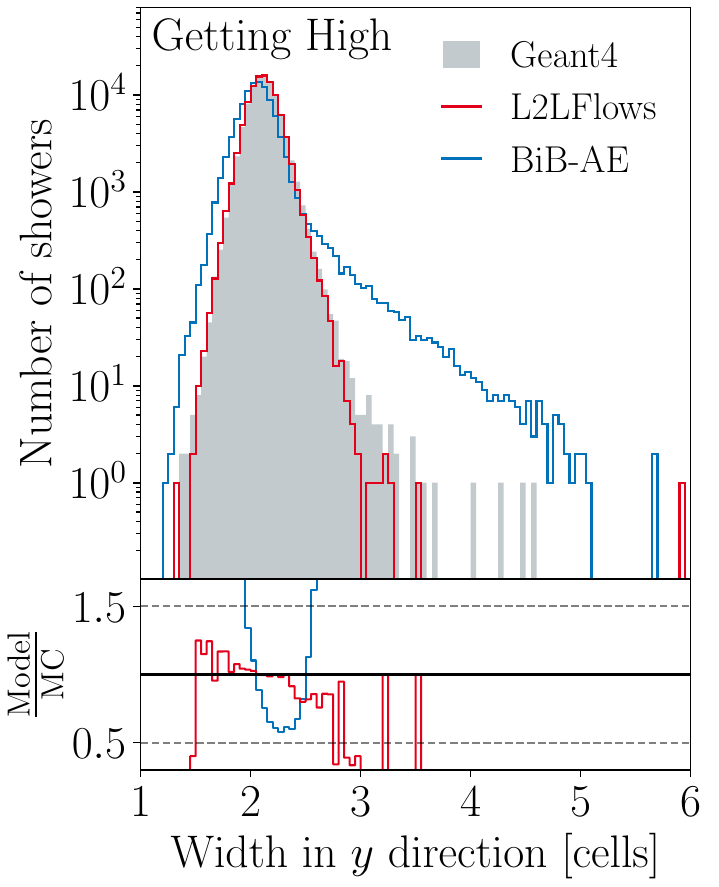}
    \caption{Standard deviation of the energy weighted distributions of longitudinal (left), \(x\) (middle), and \(y\) (right) positions evaluated on single showers and histogramed over all test showers for \textsc{GettingHigh}.}
\end{figure}
\begin{figure}[h]
    \centering
    \includegraphics[width=0.3\linewidth]{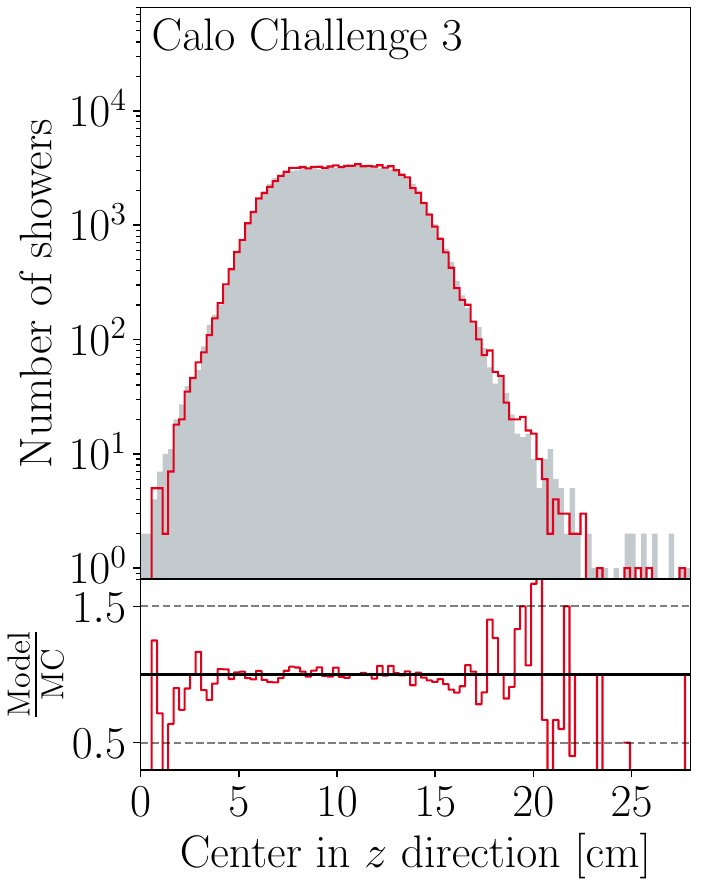}
    \includegraphics[width=0.3\linewidth]{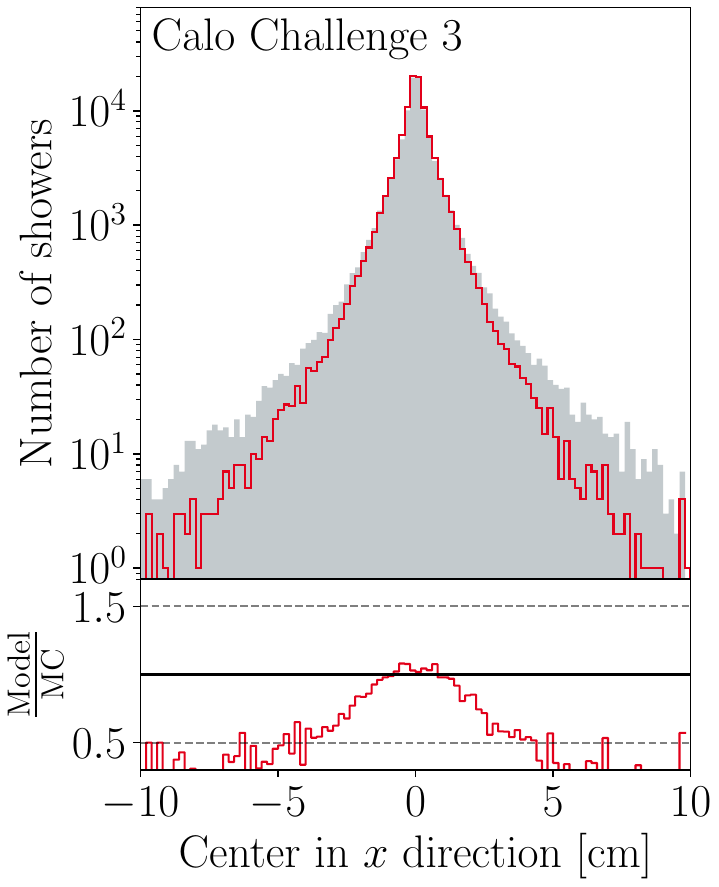}
    \includegraphics[width=0.3\linewidth]{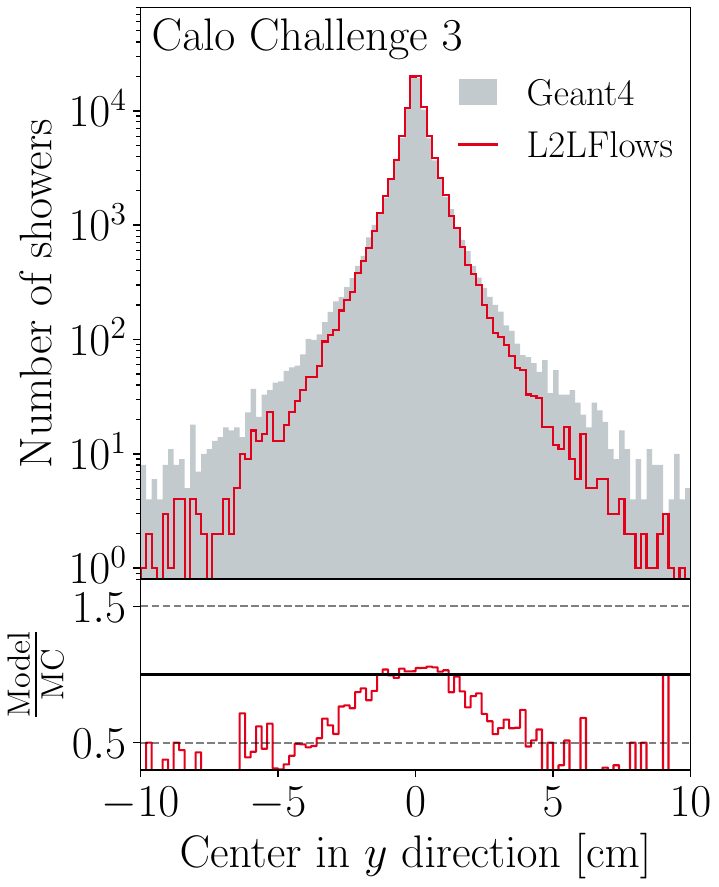}
    \caption{Center of gravity in longitudinal (left), \(x\) (middle), and \(y\) (left) direction for \textsc{CaloChallenge~3}.}
\end{figure}
\begin{figure}[h]
    \centering
    \includegraphics[width=0.3\linewidth]{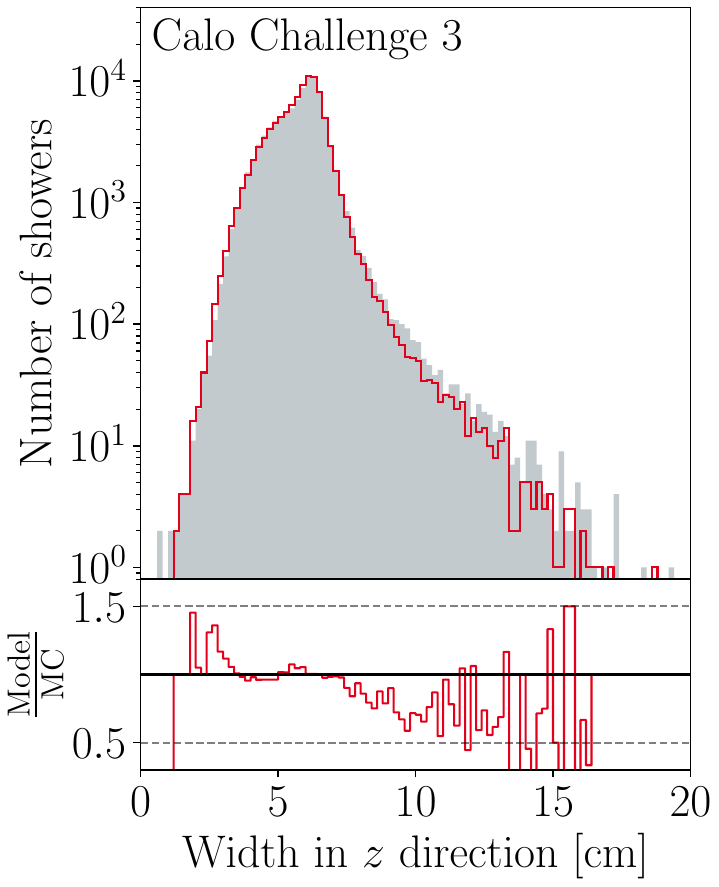}
    \includegraphics[width=0.3\linewidth]{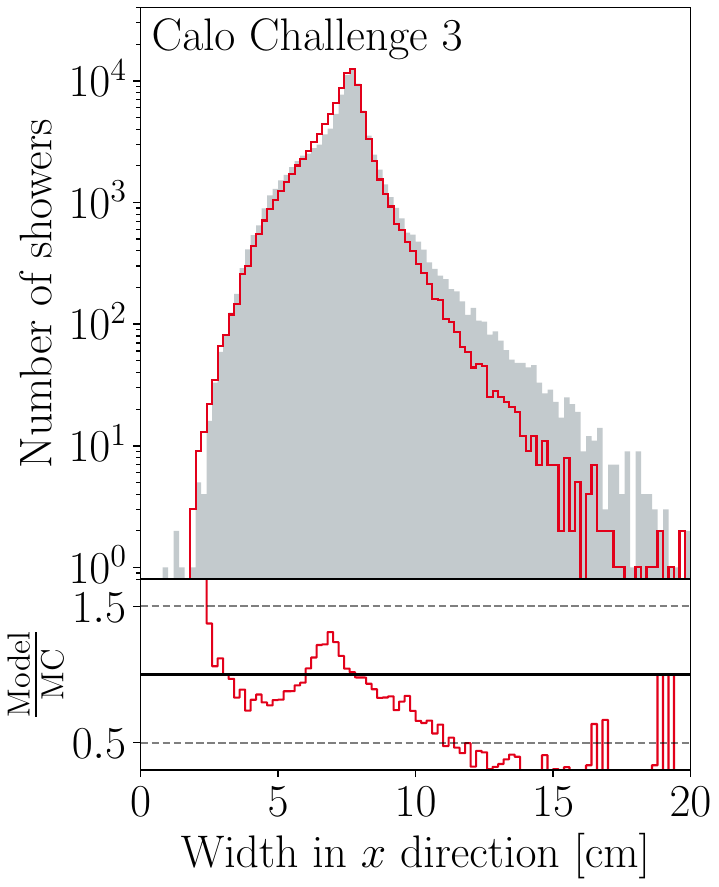}
    \includegraphics[width=0.3\linewidth]{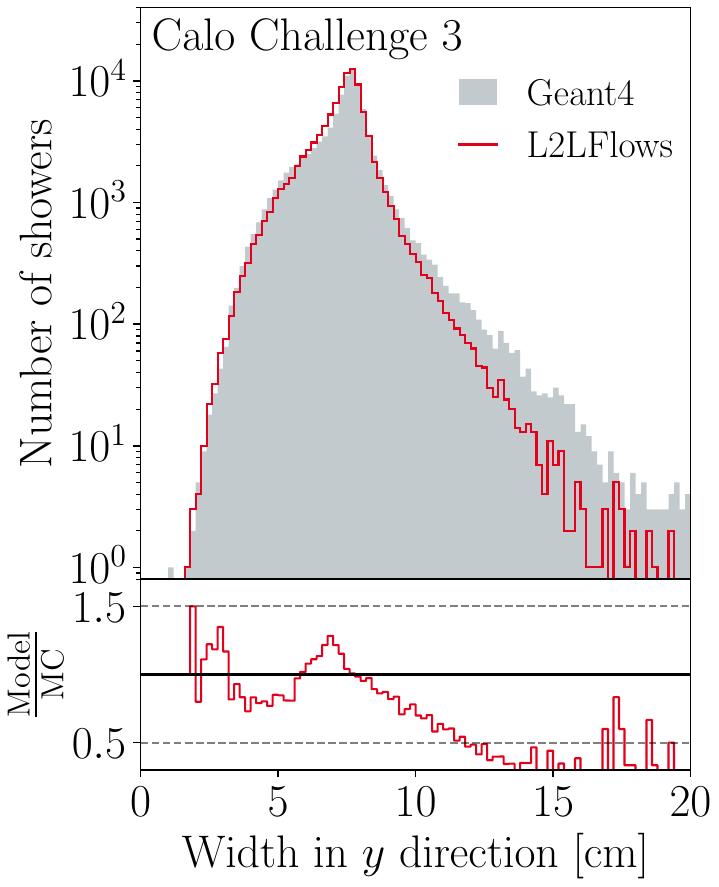}
    \caption{Standard deviation of the energy weighted distributions of longitudinal (left), \(x\) (middle), and \(y\) (right) positions evaluated on single showers and histogramed over all test showers for \textsc{CaloChallenge~3}.}
\end{figure}

\end{document}